\pdfoutput=1
\documentclass[sigconf, language=english]{acmart} %
\AtBeginDocument{%
  }

\renewcommand\footnotetextcopyrightpermission[1]{}
\acmConference[]{Preprint}
\acmDOI{}
\acmYear{}

\acmISBN{978-1-4503-XXXX-X/2018/06}

\usepackage{makecell}
\usepackage{subcaption}
\usepackage{dblfloatfix}
\usepackage{placeins}
\usepackage{setspace}
\usepackage{enumitem}
\usepackage{array}
\usepackage{multirow}
\usepackage{svg}
\usepackage{graphicx}

\renewcommand\footnotetextcopyrightpermission[1]{} %
\begin{document}

\title{SteelDB: Diagnosing Kernel-Space Bottlenecks in Cloud OLTP Databases}

\author{Mitsumasa Kondo}
\affiliation{%
  \institution{NTT, Inc.}
  \state{Tokyo}
  \country{Japan}  %
}
\email{mitsumasa.kondou@ntt.com}

\renewcommand{\shortauthors}{Kondo}

\begin{abstract}
Modern cloud OLTP databases have sought performance primarily through user-space optimization---separating storage and compute layers, or distributing transactions across multiple nodes using consensus algorithms.
This paper turns attention to a previously unexplored layer: kernel-space I/O behavior.
From an on-premises perspective, where a single server with local storage delivers excellent performance, these elaborate designs seem puzzling.
Why do cloud databases require such architectural complexity?
We investigate this through a pathological analysis of databases that rely on OS-level I/O control in cloud-specific storage environments.
We show that bottlenecks widely attributed to network or storage architectures in fact originate in kernel-space I/O behavior.
Based on this diagnosis, we derive treatment principles and realize them as SteelDB, a zero-patch architecture that improves database performance on general-purpose cloud distributed block storage through strategic I/O optimization without requiring kernel or database patches.
TPC-C evaluations demonstrate that SteelDB achieves up to 9x performance improvement at no additional cost. 
Against Amazon Aurora, SteelDB achieved 3.1x higher performance while reducing costs by 58\%, leading to a 7.3x improvement in cost efficiency.
While Aurora requires an average of 254 days for major version upgrades due to applying proprietary patches to newly released OSS databases, our zero-patch architecture reduces these software maintenance costs to near zero.
\end{abstract}

\vspace{-10pt}
\keywords{Cloud OLTP, Kernel Flusher Thread, Merge I/O, Distributed Block Storage, Zero-Patch Architecture}

\maketitle
\vspace{-5pt}
\section{Introduction}\label{sec:intro}
The database market continues to grow steadily, and by 2024, cloud databases accounted for 64\% of total market share~\cite{gartner2025dbmsmarketshare}.
As cloud adoption accelerates, the demand for high-performance yet cost-effective database solutions is a key driver of adoption.

This paper focuses on cloud databases for OLTP.
Existing cloud OLTP databases address performance challenges through two primary architectural approaches.
The first is storage-compute disaggregation databases, exemplified by Amazon Aurora~\cite{verbitskiAmazon2017} and Google AlloyDB~\cite{noauthorAlloydbNodate}. 
These systems separate compute and storage layers, transmitting only write-ahead logs from compute nodes to dedicated storage nodes where data pages are reconstructed.
The second approach is distributed transactional databases such as Google Spanner~\cite{corbettSpanner2013} and TiDB~\cite{huangTidb2020}, which distribute data across multiple nodes using consensus algorithms like Raft and Paxos to provide horizontal scalability and fault tolerance.

Both approaches incur non-trivial software complexity and higher user-facing costs.
In cloud environments, managed database services such as Amazon RDS incur approximately twice the cost of standard VM instances~\cite{noauthorAwsCalc}, primarily due to dedicated storage and management overhead.
Storage-compute disaggregated databases such as Aurora further increase costs to roughly three times that of standard instances, reflecting the additional expense of proprietary storage systems.
Beyond infrastructure costs, these systems require substantial engineering effort to maintain compatibility with upstream open-source releases.
Distributed databases incur overhead from transaction coordination and face challenges with common SQL patterns such as distributed JOINs.

A fundamental question remains: why do cloud databases require such elaborate architectures?
In on-premises environments, a single server with local storage delivers excellent performance using standard databases like PostgreSQL~\cite{groupPostgresql2024}.
What makes cloud environments so different?
Prior research~\cite{verbitskiAmazon2017,pangUnderstanding2024} has attributed cloud database performance challenges primarily to network limitations---specifically, packets per second (PPS) and bandwidth constraints when accessing distributed storage. 
Storage-compute disaggregation databases were designed to address these bottlenecks.
However, modern cloud infrastructure has evolved significantly.
VM instances now offer network bandwidth of 10--100 Gbps, and core networking equipment achieves 100--400 Gbps.
If network performance is no longer the primary limiting factor, what is the fundamental bottleneck?

This paper makes two contributions.
First, we conduct a pathological analysis of databases that rely on OS-level I/O control in cloud-specific storage environments and identify kernel-space bottlenecks that have been overlooked in previous research.
We demonstrate that performance limitations widely attributed to network or storage architectures instead 
originate in the suboptimal interaction between buffered I/O mechanisms and high-latency cloud distributed block storage.
Second, based on this diagnosis, we derive treatment principles and realize them as SteelDB, a zero-patch
architecture that improves performance through strategic I/O optimization without requiring kernel or database patches.
TPC-C evaluations demonstrate that SteelDB achieves up to 9x performance improvement at no additional cost.
Against Amazon Aurora, SteelDB achieved 3.1x higher performance while reducing costs by 58\%, leading to a 7.3x improvement in cost efficiency.

The paper is structured as follows.
Section 2 lays the groundwork, describing how database I/O, distributed block storage, and kernel I/O control mechanisms operate under normal conditions.
Section 3 presents our pathological examination, documenting symptoms of performance degradation through controlled experiments on distributed storage.
Section 4 diagnoses the root causes.
Section 5 derives treatment principles and describes the SteelDB architecture as their realization.
Section 6 evaluates treatment effectiveness through TPC-C benchmarks and comparisons with commercial cloud databases.
Section 7 discusses related work, and Section 8 concludes.

\section{TECHNICAL BACKGROUND}\label{sec:techback}
This section describes the normal operation of database I/O, distributed storage, and kernel I/O control mechanisms---the foundation for our pathological analysis in subsequent sections.

\vspace{-3pt}
\subsection{Database}\label{subsec:db}
\subsubsection{Database I/O Control}\label{subsubsec:dbiocont}
Database I/O control methods are primarily classified into database-controlled and OS-controlled approaches.
In the database-controlled approach, the database manages I/O internally rather than relying on OS-level buffered I/O.
This typically involves direct I/O for local storage access, but also encompasses architectures that bypass local storage entirely---such as Aurora's WAL-shipping to dedicated storage nodes. 
Storage-compute disaggregated databases such as Aurora and AlloyDB, as well as Oracle Database~\cite{noauthorOracleNodate}, use this method. 
However, dedicating development resources to I/O control alone is impractical when many database features remain to be implemented.
Many databases therefore delegate I/O control to the OS~\cite{qianCombining2024, kimRevitalizing2023}.

OS-controlled databases such as PostgreSQL and RocksDB~\cite{dongRocksdb2021} rely on the OS's buffered I/O by default, delegating the costly I/O control implementation to the kernel and concentrating development resources on database-specific features. 
This design embodies the UNIX philosophy~\cite{macilroy} that "Make each program do one thing well" and "Small is beautiful"---a strategy that trades I/O control for portability across platforms.
While direct I/O can outperform buffered I/O when implemented with sufficient sophistication, its adoption barrier is high; buffered I/O remains prevalent even in HPC environments~\cite{qianCombining2024}. 
Following this principle, PostgreSQL, with over 20 years of history, prioritizes database feature development by delegating I/O control to the OS, making it one of the most standards-compliant SQL databases in the world~\cite{celkoJoe2010}. MySQL~\cite{noauthorMysqlNodate} adopts a hybrid approach that supplements database-side direct I/O with the OS's buffered I/O.

Delegating I/O to the OS can also be efficient: buffered I/O avoids the per-operation system call overhead of user-space direct I/O.
This paper focuses on databases that delegate I/O control to the operating system, using PostgreSQL as a representative example.
PostgreSQL performs I/O operations in 8 KB pages, including key data structures such as WAL, tables, and indexes. I/O system calls are issued in 8 KB units for both random and sequential I/O. While WAL performs sequential writes, table and index access patterns vary between sequential and random I/O depending on the SQL operation.
This I/O pattern can be observed during PostgreSQL operations by running the biopattern tool from the BCC~\cite{noauthorBccNodate} (BPF Compiler Collection) tool, which utilizes eBPF~\cite{noauthorEbpfNodate} (extended Berkeley Packet Filter).

\vspace{-5pt}
\subsubsection{Write-Ahead Logging. }\label{subsubsec:wal}
WAL~\cite{mohanAries1992} ensures data durability by recording changes sequentially before applying them to data files.
In PostgreSQL, WAL records are stored in 8 KB pages, and writes are performed using buffered I/O at each transaction commit, followed by an fsync operation to ensure persistence.
The WAL writer process can incur more than 100 fsync operations per second, making it a dominant factor in write performance.
Since WAL writes are processed synchronously, lower storage latency directly results in better write performance.

\vspace{-5pt}
\subsubsection{Tables, Indexes, and Tablespaces. }\label{subsubsec:tablespace}
Data recorded in WAL is stored in shared buffers and written to disk during checkpoint operations or when buffer space needs to be freed, together with index data.
These disk writes use buffered I/O: data is first written to memory as a dirty page and then flushed to disk by the OS's periodic flush operations or by the final fsync at checkpoint completion.
Writes except for this final fsync are handled asynchronously.
Tablespaces enable users to allocate tables and indexes to specific storage locations, distributing data across multiple disks.
Once common for I/O parallelism, tablespaces fell into disuse with the advent of hardware RAID and high-performance storage.
SteelDB revives this standard feature for a different purpose.%

\vspace{-5pt}
\subsection{Distributed Block Storage}\label{subsec:disblock}
Cloud databases rely on network-attached distributed block storage for persistence.
In cloud environments, local storage is typically volatile---VMs may migrate across host servers during restarts or rebalancing---making distributed block storage the standard practice for databases. While locally attached NVMe storage is available on some instance types, data stored on these devices is lost when the instance is stopped or restarted, making them unsuitable for database use cases that require durability.
While distributed storage encompasses both object storage (e.g., Amazon S3~\cite{warfieldAmazon2023}) and block storage (e.g., Amazon EBS~\cite{noauthorCloudNodate}), this paper focuses on distributed block storage.
Since major cloud providers disclose only partial storage internals (e.g., EBS's Physalia~\cite{brooker2020millions}) and not data-path details such as QoS enforcement, we use Ceph~\cite{aghayevFile2019, weilCeph2006}, a widely used open-source implementation, as a representative example.
These systems ensure high availability and data consistency through distributed consensus algorithms such as Raft~\cite{ongaroSearch2014} and Paxos~\cite{lamportPartTime1998}, replicating data across multiple servers and disks.
In Ceph, the CRUSH algorithm distributes data via Placement Groups across multiple physical disks, achieving RAID-like throughput.

While distributed block storage appears as a standard block device to users, its performance profile differs sharply from local storage.
Distributed storage is designed to provide stable I/O to multiple concurrent users~\cite{gulatiDClock2007} through QoS mechanisms such as mClock~\cite{gulatiMclock2010}, rather than allowing a single user to fully exploit the underlying hardware.
Additionally, performing synchronous I/O operations across multiple disks over a network results in high I/O latency---
particularly harmful to small I/O operations, where per-operation latency dominates throughput.
In contrast, local storage accesses disks directly via internal buses (e.g., SATA, NVMe), achieving extremely low latency and high throughput even with small I/O sizes.
On-premises deployments further benefit from hardware RAID with battery-backed write cache---functionality unavailable in cloud distributed storage, where only software RAID is offered.
These gaps underlie the on-premises performance advantage and motivate the examination in Section 3.

\subsection{Kernel I/O Control}\label{subsec:kernelio}
This section explains the disk I/O features provided by the kernel, focusing on the Linux kernel~\cite{torvaldsTorvaldslinux2025}, which is widely used in many database systems.
The Linux kernel primarily offers two types of disk I/O interfaces: buffered I/O and direct I/O.
Buffered I/O is the standard method for processing file system data and is widely adopted by major databases such as PostgreSQL, MySQL, and RocksDB~\cite{dongRocksdb2021}.
It often outperforms direct I/O because the kernel applies its own I/O optimizations transparently.
As noted in Section 2.1, most databases rely on buffered I/O.

\subsubsection{Dirty Page, and Write-Back. }\label{subsubsec:page-cache}
\begin{figure}[t]
\centering
\includegraphics[width=0.95\linewidth]{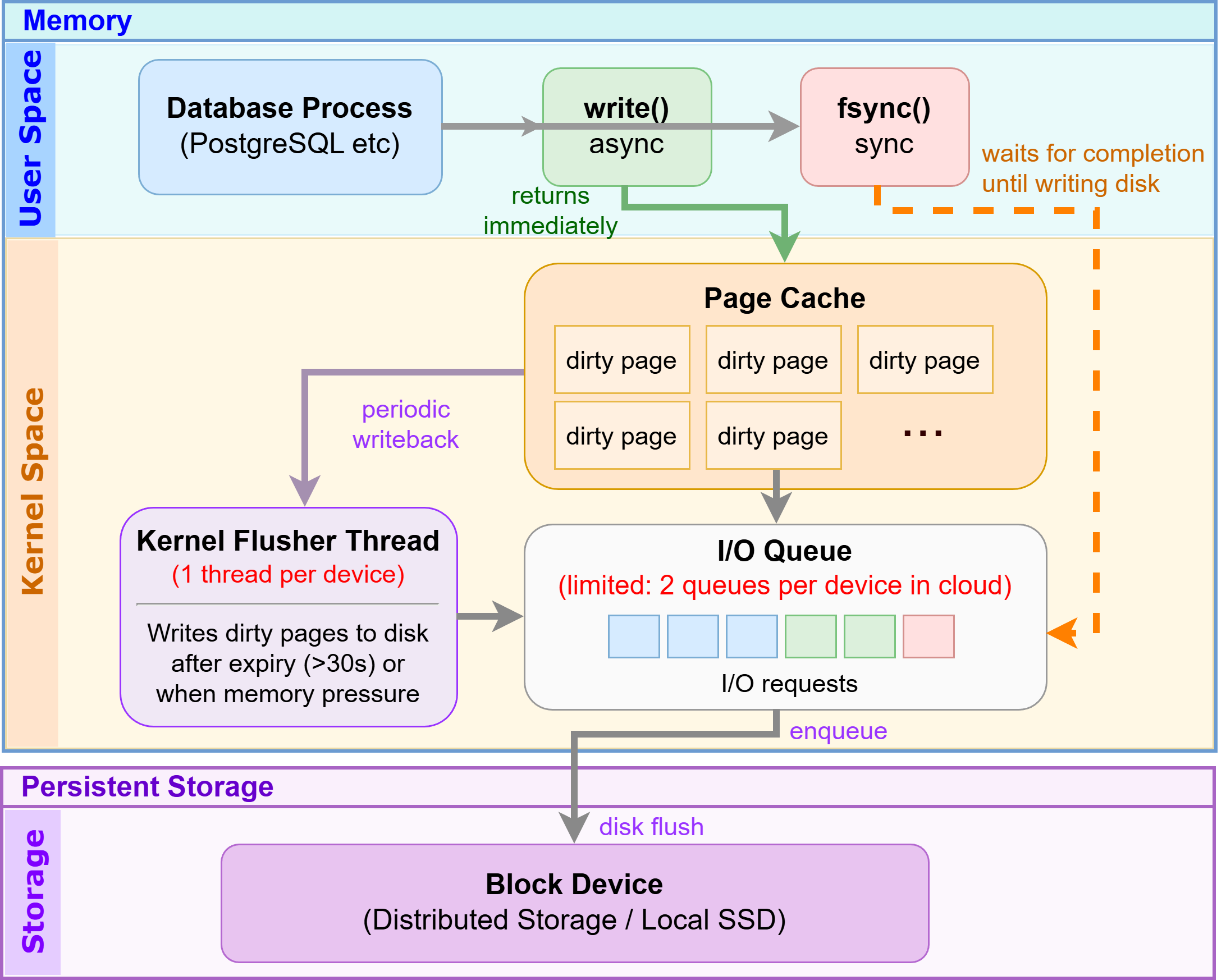}
\vspace{-10pt}
\caption{Linux Kernel Writeback Architecture} 
\vspace{-9pt}
\label{fig:linux-writeback}
\end{figure}

In buffered I/O write operations, data is not immediately written to disk but stored in the kernel page cache as a dirty page.
These dirty pages are periodically flushed to disk by the Kernel Flusher Thread (KFT)---a process known as write-back.
Write-back batches multiple writes and dynamically adjusts its behavior based on system load and memory usage.
For synchronous disk writes, the fsync system call flushes dirty pages and waits for completion, ensuring data durability.
The Linux kernel writeback architecture is shown in Figure \ref{fig:linux-writeback}.

\subsubsection{Kernel Flusher Thread}\label{subsubsec:KFT}
The Kernel Flusher Thread (KFT) is a background process in the OS kernel that periodically flushes dirty pages from memory to disk.
Dirty pages are generated by write system calls and remain in memory until KFT writes them back to persistent storage.
KFT is also responsible for handling disk writes triggered by the fsync system call.
Typically, one KFT is assigned per block device; even when multiple physical disks are configured as a RAID array, a single KFT handles the entire array.
These characteristics become particularly relevant in cloud environments, as discussed in subsequent sections.
KFT is visible in process listings (e.g., via top or ps) as kworker/<work queue>+flush-259:0, where 259:0 is the identifier of the target block device.

\subsubsection{I/O-less Dirty Throttling. }\label{subsubsec:ioless-dirty-throttling}
I/O-less Dirty Throttling~\cite{fengguangFengguangNodate} is an OS feature introduced in Linux kernel version 3.2 to prevent memory exhaustion by gradually limiting the speed of the write system call when the number of dirty pages continuously exceeds a certain threshold.
While essential for system stability, this mechanism can severely degrade database performance on high-latency storage.
By throttling the frequency of write system calls, it ensures that OS processes can operate stably without halting I/O operations.
However, if the flushing of dirty pages to disk remains insufficient, the OS may completely halt the execution of write system calls.
This mechanism, while essential for system stability, can significantly impact database performance when storage latency is high.

\subsubsection{I/O Queues}\label{subsubsec:ioqueues}

The kernel's I/O queues are essential for efficiently processing I/O requests for both disk and network storage.
For high-speed devices like NVMe, Linux uses the Block Layer Multi-Queue (blk-mq) framework, which enables parallel processing of multiple I/O requests, improving both throughput and latency.
Among available scheduling algorithms, the none scheduler processes I/O requests in arrival order without reordering, achieving low-latency disk I/O with minimal overhead while supporting Merge I/O.
In cloud environments, the none scheduler is recommended; all experiments in this paper were conducted using it.
The total number of Linux multi-queues is determined by the minimum of vCPU core count and device queue count.
Critically, if the device exposes fewer queues than available vCPU cores, users cannot increase the queue count through any configuration---this is a hardware-level constraint imposed by the block device.
Each queue is assigned to a specific vCPU, and I/O commands executed on that vCPU are processed through its corresponding queue.
This mechanism also applies when multiple disks are present, with each disk receiving its own set of queue assignments.
The number of I/O queues can be inspected via /sys/block/<device name>/mq.

\begin{figure*}[t]
    \centering
    \vspace{-0.3cm} %
    
    \begin{subfigure}{0.247\textwidth}
        \centering
        \includegraphics[trim=30 87 30 50,width=\linewidth]{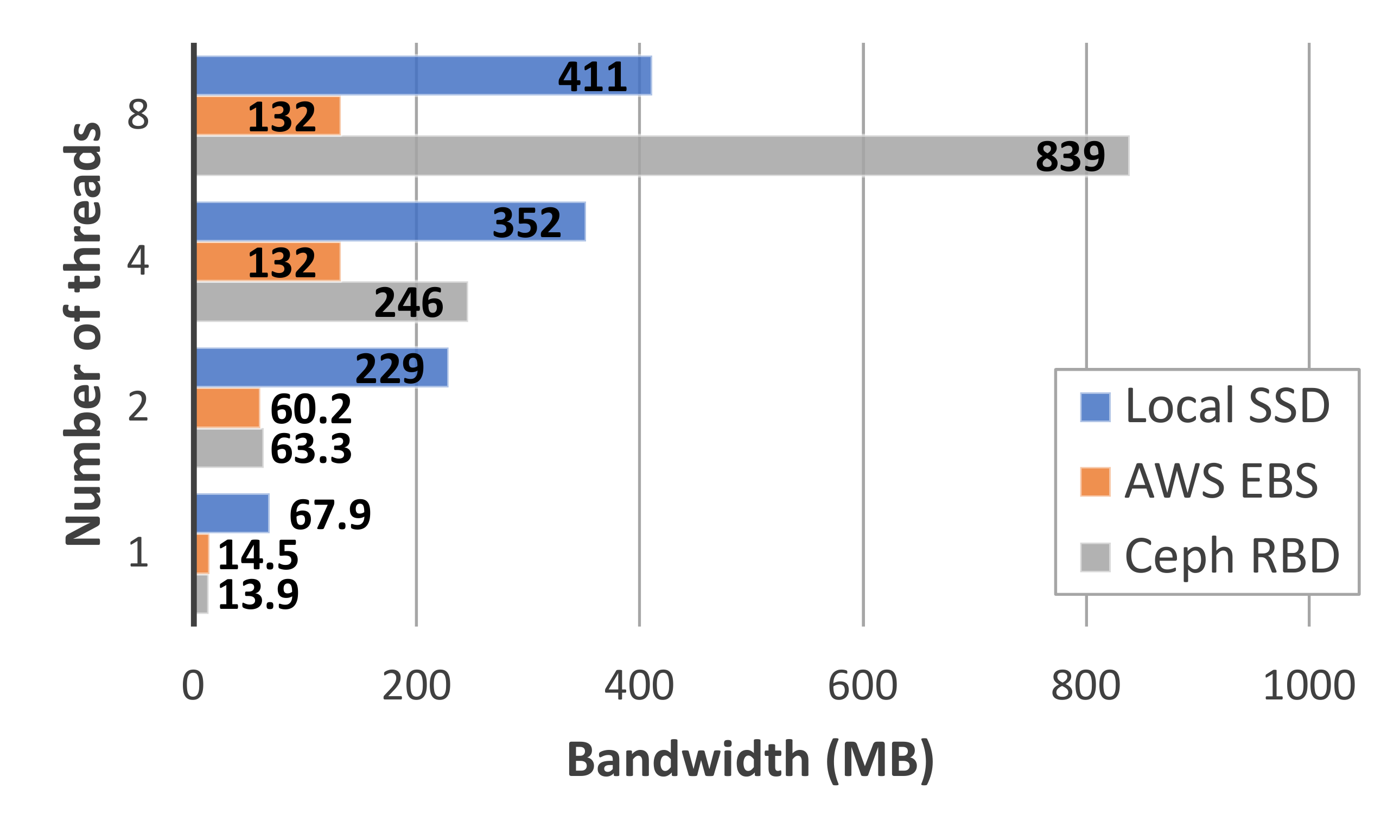}
        \caption{Random Read}
     \end{subfigure}
    \begin{subfigure}{0.247\textwidth}
        \centering
        \includegraphics[trim=25 87 30 50,width=\linewidth]{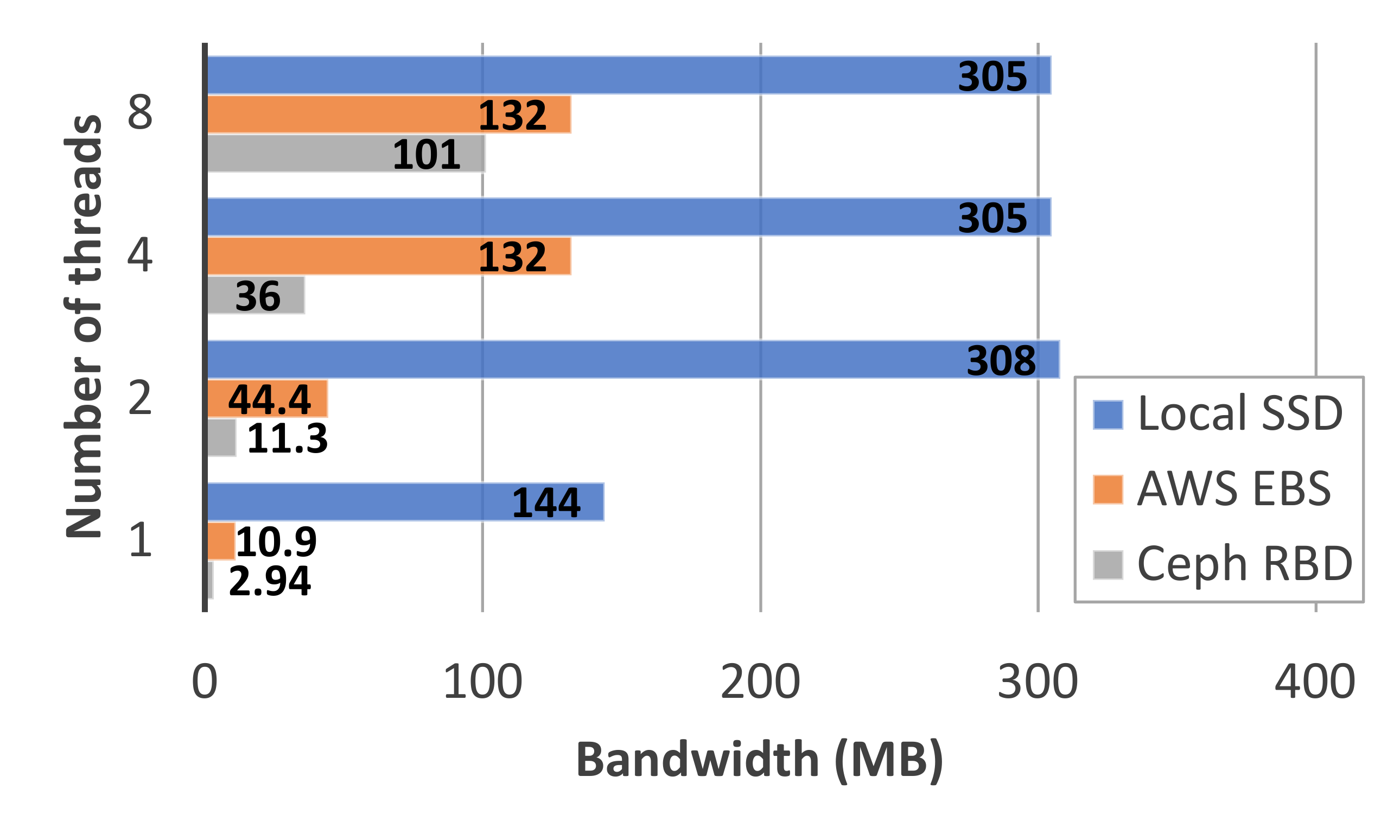}
        \caption{Random Write}
     \end{subfigure}
    \begin{subfigure}{0.247\textwidth}
        \centering
        \includegraphics[trim=25 87 30 50,width=\linewidth]{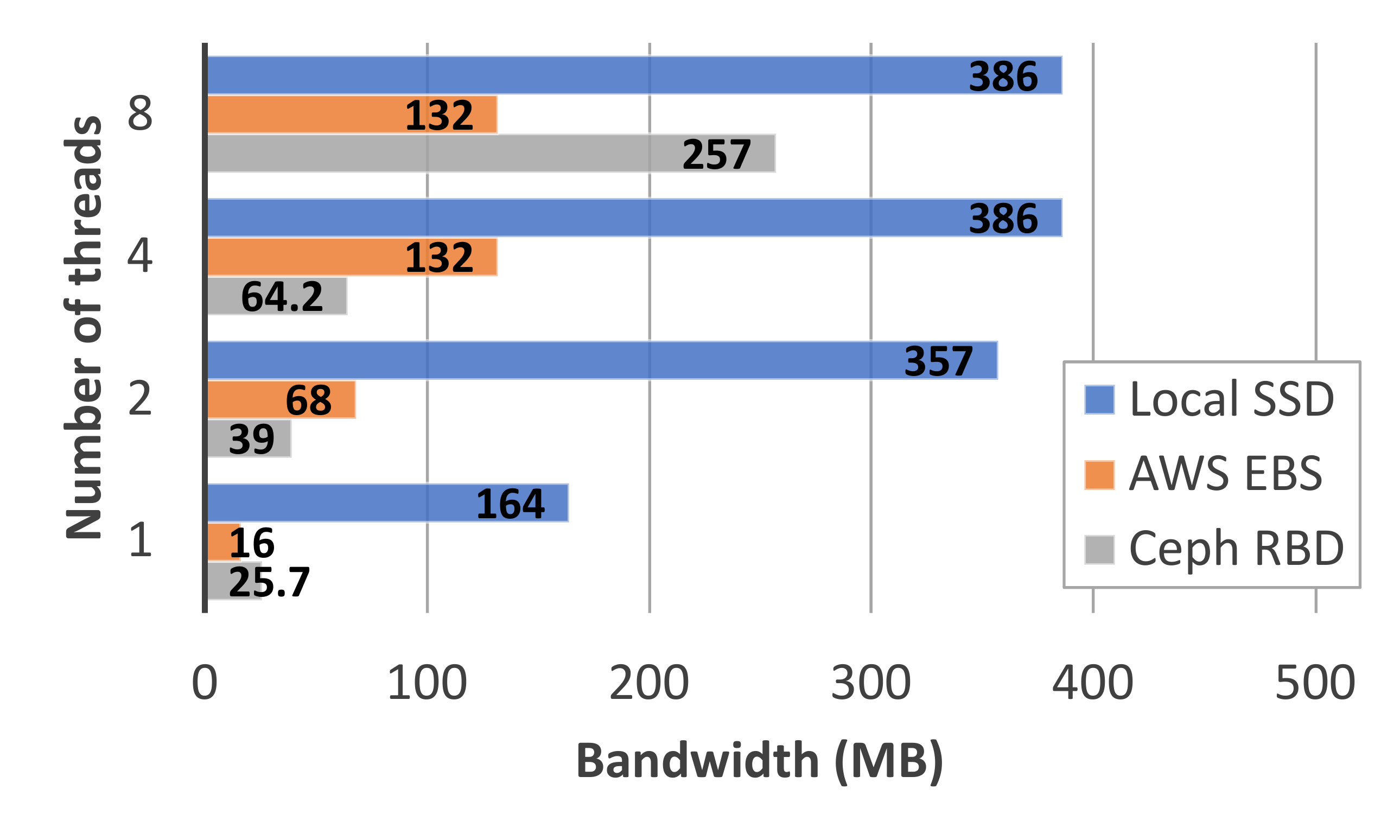}
        \caption{Sequential Read}
     \end{subfigure}
    \begin{subfigure}{0.246\textwidth}
        \centering
        \includegraphics[trim=25 87 30 50,width=\linewidth]{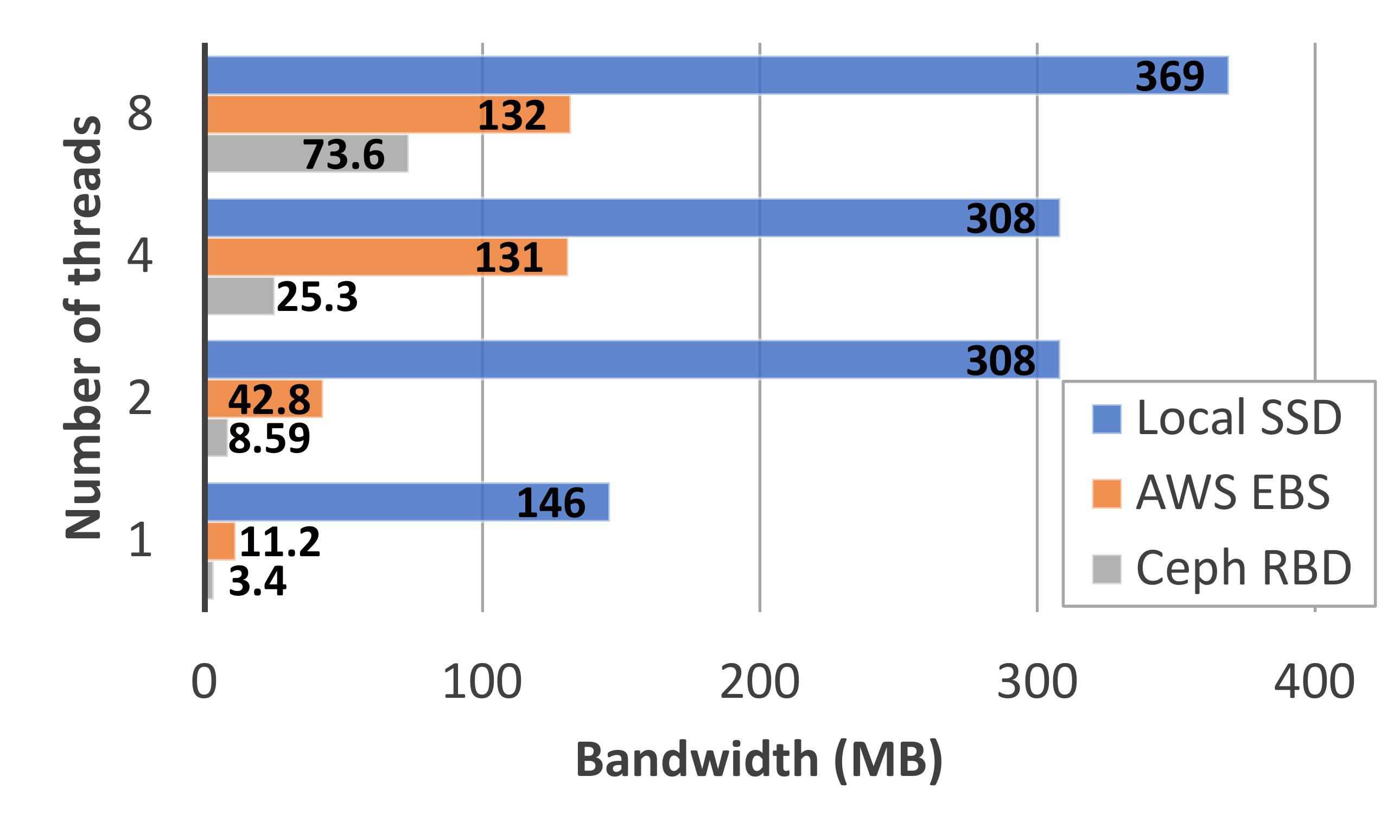}
        \caption{Sequential Write}
     \end{subfigure}
    \vspace{-23pt}
    \caption{Direct I/O Bandwidth Measurement with Varying Thread Counts}
    \vspace{-6pt}
    \label{fig:dio-test-threads}
\end{figure*}

\begin{figure*}[t]
    \centering
    
    \begin{subfigure}{0.247\textwidth}
        \centering
        \includegraphics[trim=30 87 30 50,width=\linewidth]{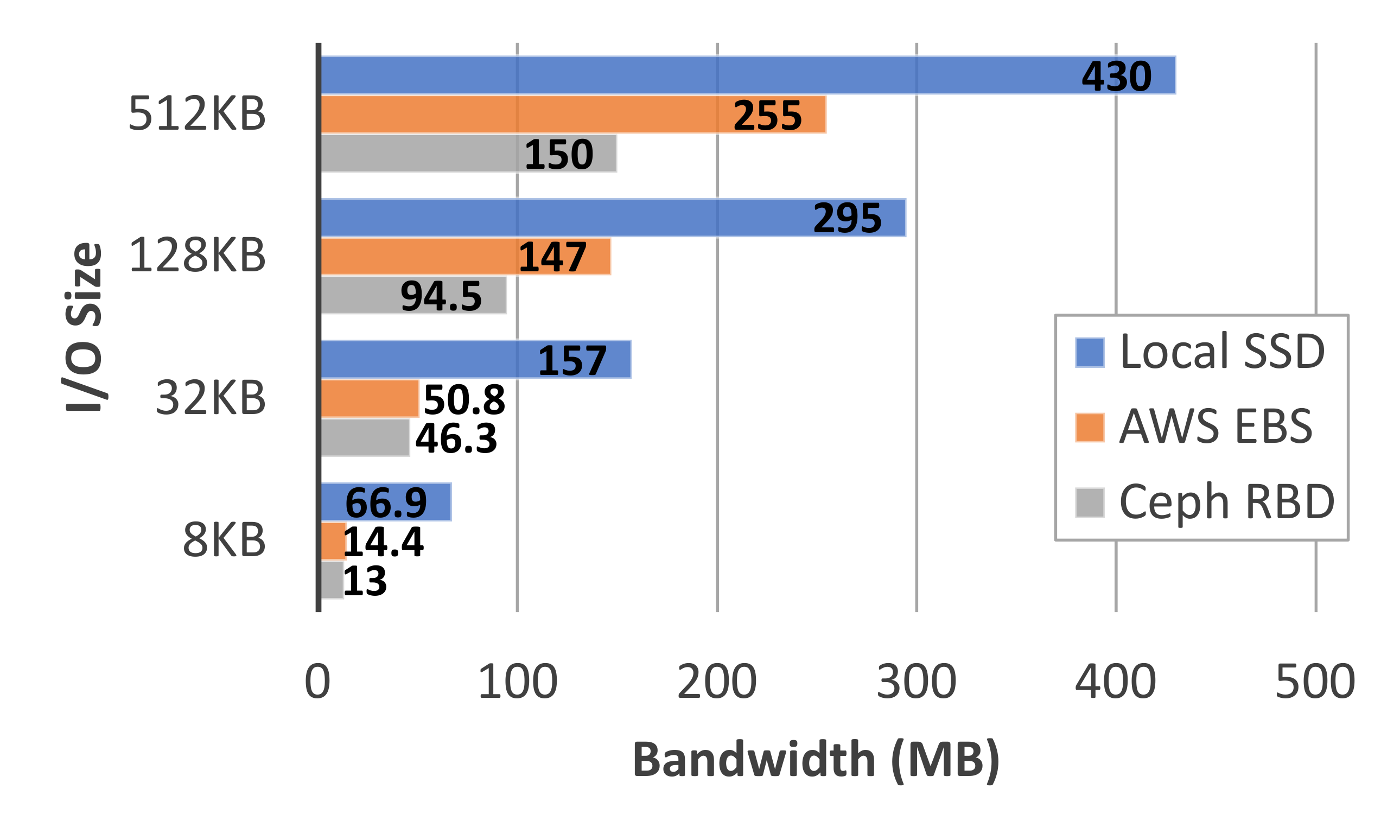}
        \caption{Random Read}
     \end{subfigure}
    \begin{subfigure}{0.247\textwidth}
        \centering
        \includegraphics[trim=25 87 30 50,width=\linewidth]{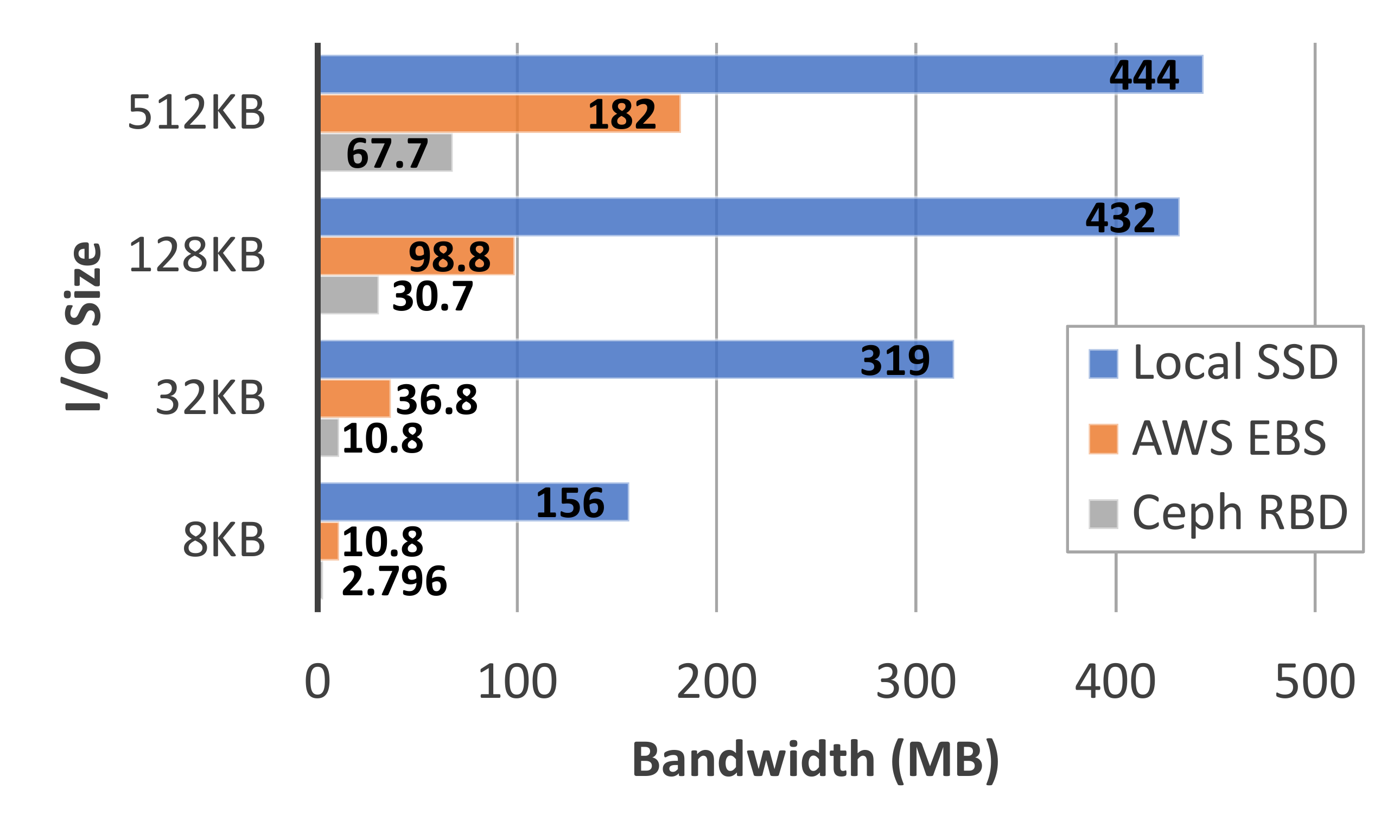}
        \caption{Random Write}
     \end{subfigure}
    \begin{subfigure}{0.247\textwidth}
        \centering
        \includegraphics[trim=25 87 30 50,width=\linewidth]{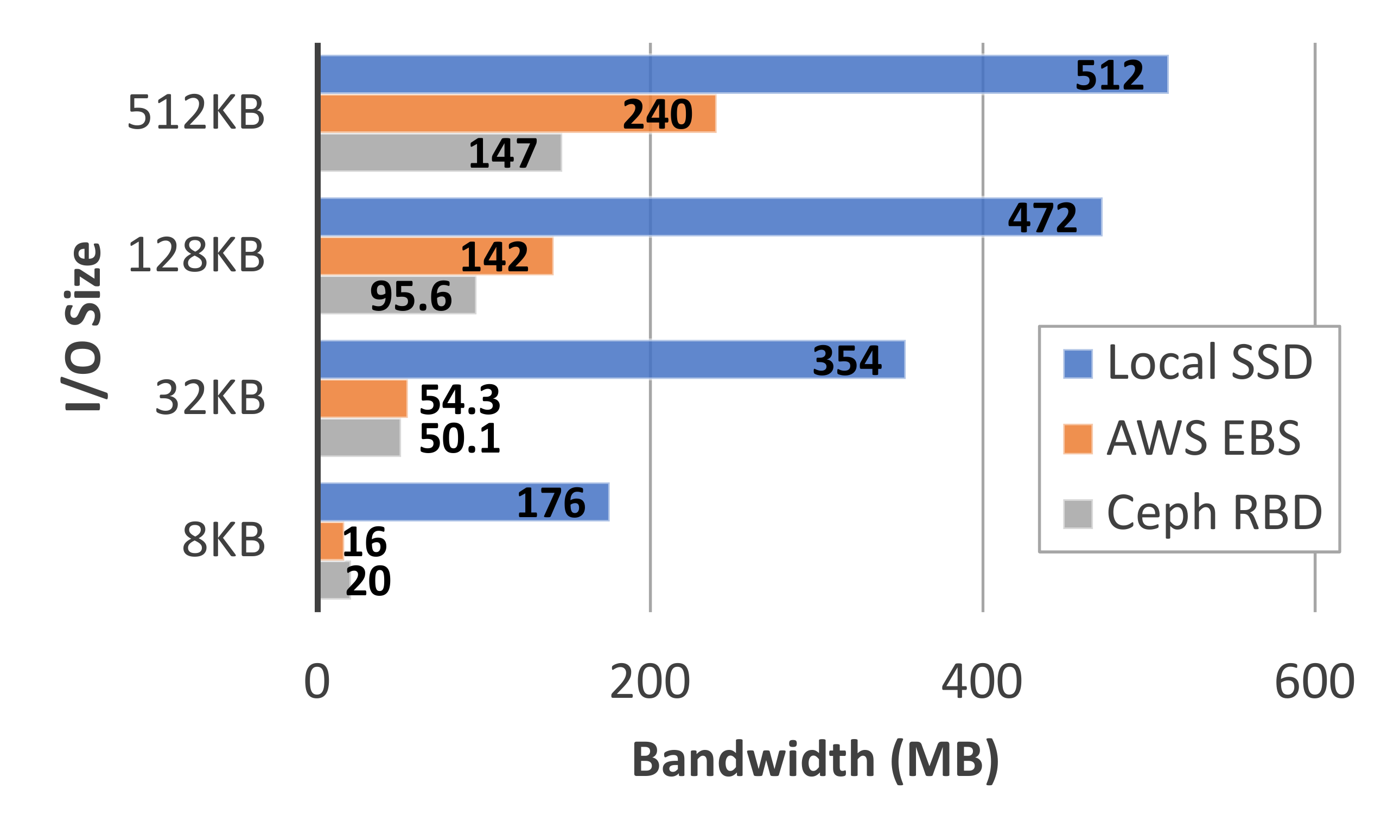}
        \caption{Sequential Read}
    \end{subfigure}
    \begin{subfigure}{0.247\textwidth}
        \centering
        \includegraphics[trim=25 87 30 50,width=\linewidth]{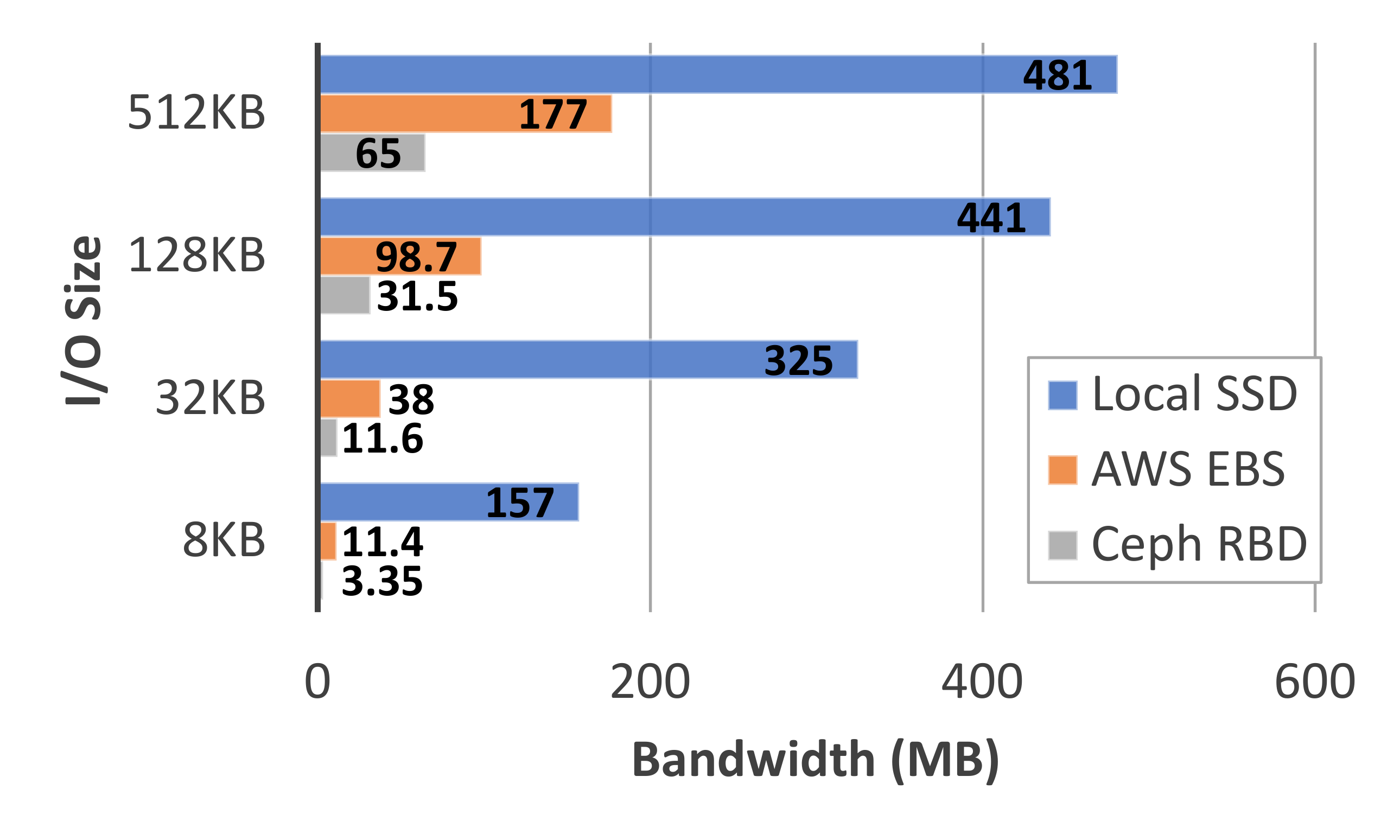}
        \caption{Sequential Write}
     \end{subfigure}

\vspace{-12pt}
    \caption{Direct I/O Bandwidth Measurement with Varying I/O Size}
    \label{fig:dio-test-iosize}
    \vspace{-5pt}
\end{figure*}

\begin{figure}[ht]
  \vspace{-7pt}
  \centering
  \includegraphics[trim=0 57 0 110,width=0.825\linewidth]{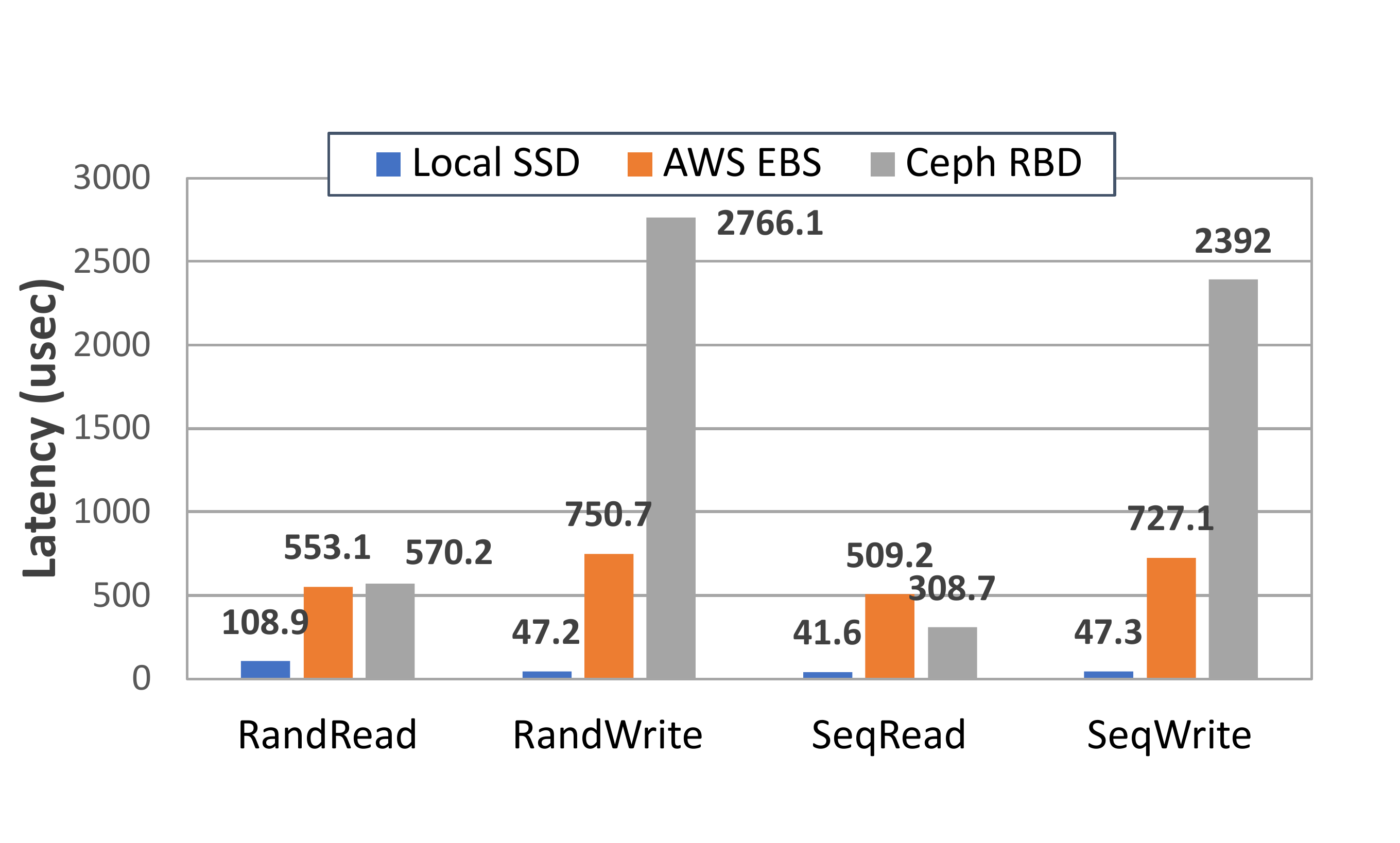}
  \vspace{-15pt}
  \caption{Direct I/O Latency Measurement with 8 KB I/O Size and a Single Thread} 
  \label{fig:dio-lat-single-8kb}
  \vspace{-7pt}
\end{figure}

\begin{figure}[ht]
    \centering
    \fbox{%
        \scriptsize
        \setlength{\tabcolsep}{1pt}
        \begin{tabular}{@{}ll@{\hspace{4pt}}ll@{}}
        \textbf{Server:} & ProLiant DL325 Gen10 Plus x6 &
        \textbf{Disk(OS):} & SATA SSD 1TB \\
        \textbf{CPU:} & EPYC 7302P 16C/32T(3GHz) &
        \textbf{Disk(Ceph):} & SATA SSD 1TB x 7 \\
        \textbf{Memory:} & DDR4 192GB &
        \textbf{Network:} & Mellanox ConnectX-6 DX 100Gbps x2 \\
        \textbf{OS:} & Rocky Linux 9.3 (x86\_64) &
        \textbf{Switch:} & Edgecore AS7726-32X (SONiC) \\
        \end{tabular}
    }
    \vspace{-12pt}
    \caption{Specifications of a Ceph Cluster}
    \label{fig:ceph_servers}
    \vspace{-8pt}
\end{figure}

\subsubsection{Merge I/O}\label{subsubsec:mergeio}
Merge I/O is a feature in the Linux kernel that combines multiple read or write requests into a single larger request during buffered I/O processing.
First implemented in Linux kernel version 1.0 (1994) to optimize I/O for head-driven storage devices such as HDDs, it remains effective in modern kernels.
Aggregating multiple I/O operations increases the size of each I/O command, improving disk I/O efficiency.
For example, if 30 write requests of 8 KB each are issued to consecutive blocks, Merge Write consolidates them into a single 240 KB write operation.
This reduces IOPS consumption by up to 30×.
This feature is particularly beneficial in cloud environments with distributed block storage, where higher IOPS settings lead to increased costs.
Merge I/O thus improves throughput while reducing IOPS consumption---a dual benefit of performance and cost in cloud environments where IOPS are billed.
Furthermore, merge I/O can be applied even to I/O commands from different processes, as long as the target blocks are adjacent.
For merge I/O to be executed, the I/O commands must reside in the same I/O queue and target contiguous blocks.

Proactive exploitation of merge I/O is central to our approach---rather than relying on merge I/O as a passive kernel optimization, SteelDB creates conditions that maximize its effectiveness.

\subsubsection{Summary of Kernel I/O Control}\label{subsubsec:2summary}
Under normal conditions---low-latency local storage with sufficient I/O queue capacity---these kernel mechanisms work harmoniously. 
The flusher thread keeps pace with dirty page generation, I/O queues process requests efficiently, and merge I/O consolidates sequential writes.
However, as we demonstrate in Section 3, cloud distributed block storage disrupts these assumptions.

\section{PATHOLOGICAL EXAMINATION}\label{sec:pre-exp}
This section examines the I/O behavior of cloud distributed block storage to identify performance pathologies under database-like workloads.
Using local storage as a healthy baseline, we identify conditions that exacerbate or relieve these symptoms.
Here, we refer to ``healthy'' behavior in the sense that it reflects the performance characteristics implicitly assumed by typical OS I/O heuristics, rather than making a normative claim about storage systems.

\subsection{Examination Setup}\label{subsec:pre-exp-settp}
We compare two distributed storage systems---AWS EBS (gp3) and Ceph RBD (RADOS Block Device)---against a local SATA SSD as a healthy baseline.
All configurations used the ext4 file system.
The specifications of the Ceph cluster are shown in Figure~\ref{fig:ceph_servers}.
All experiments in this paper were conducted using Rocky Linux 9.3 with Linux kernel version 5.14.
The EBS configuration set the IOPS limit to 16,000 and the bandwidth limit to 1,000 MB/s.
For Ceph's RBD, the block size on the Ceph side was configured to 1 MB, and the Ceph version used was 17.2.4.
We used fio~\cite{axboeFlexible2022} with the io\_uring engine, enabling both direct I/O (D-IO) and buffered I/O (B-IO) measurements in the same environment.
I/O operations were conducted on a 200 GB file pre-created with the dd command, measured over 300 seconds.

\begin{table*}[t]
    \centering
    \caption{Buffered I/O Performance Analysis Using fio and iostat with 8 KB I/O Size and 8 Threads}
    \vspace{-17pt}    
    \label{tab:buf_fio}

    \renewcommand{\arraystretch}{1.5} %
    \begin{tabular}{ccc} %
    \multicolumn{1}{c}{\small \textbf{(a) Local SSD Results}} & \multicolumn{1}{c}{\small \textbf{(b) AWS EBS Results}} & \multicolumn{1}{c}{\small \textbf{(c) Ceph RBD Results}} \\ %
        \hspace{-0.45cm}
        \renewcommand{\arraystretch}{1.1} %
        \resizebox{5.9cm}{!}{
        \huge
        \begin{tabular}{lccc} %
        \toprule
        & \large{D-IO (MB/s)} & \large{B-IO (MB/s)} & \large{B-IO Size Ave (KB)} \\
        \midrule
        \large{\textbf{RandRead}}  & \LARGE{411} & \LARGE{400} & \LARGE{8.005} \\
        \large{\textbf{RandWrite}} & \LARGE{305} & \LARGE{251} & \LARGE{8.819} \\
        \large{\textbf{SeqRead}}   & \LARGE{386} & \LARGE{539} & \LARGE{878.9} \\
        \large{\textbf{SeqWrite}}  & \LARGE{369} & \LARGE{343} & \LARGE{998.1} \\
        \bottomrule
        \end{tabular}
        }\hspace{-0.24cm}
        &
        \hspace{-0.24cm}
        \renewcommand{\arraystretch}{1.1} %
        \resizebox{5.9cm}{!}{
        \huge
        \begin{tabular}{lccc} %
        \toprule
        & \large{D-IO (MB/s)} & \large{B-IO (MB/s)} & \large{B-IO Size Ave (KB)} \\
        \midrule
        \large{\textbf{RandRead}}  & \LARGE{132} & \LARGE{128} & \LARGE{8.000} \\
        \large{\textbf{RandWrite}} & \LARGE{132} & \LARGE{144} & \LARGE{8.948} \\
        \large{\textbf{SeqRead}}   & \LARGE{132} & \LARGE{215} & \LARGE{170.5} \\
        \large{\textbf{SeqWrite}}  & \LARGE{132} & \LARGE{487} & \LARGE{122.0} \\
        \bottomrule
        \end{tabular}
        }\hspace{-0.24cm}
        &
        \hspace{-0.24cm}
        \renewcommand{\arraystretch}{1.1} %
        \resizebox{5.9cm}{!}{
        \huge
        \begin{tabular}{lccc} %
        \toprule
        & \large{D-IO (MB/s)} & \large{B-IO (MB/s)} & \large{B-IO Size Ave (KB)} \\
        \midrule
        \large{\textbf{RandRead}}  & \LARGE{839}  & \LARGE{614}  & \LARGE{8.000} \\
        \large{\textbf{RandWrite}} & \LARGE{101}  & \LARGE{51.1} & \LARGE{8.395} \\
        \large{\textbf{SeqRead}}   & \LARGE{257}  & \LARGE{131}  & \LARGE{113.9} \\
        \large{\textbf{SeqWrite}}  & \LARGE{73.6} & \LARGE{225}  & \LARGE{327.0} \\
        \bottomrule
        \end{tabular}
        }\hspace{-0.3cm}
        \\
        
    \end{tabular}
    \vspace{-8pt}
\end{table*}

\subsection{Stress Conditions}
We identify conditions that exacerbate symptoms.
As a baseline observation, distributed storage exhibits latencies more than 
10x higher than local SSDs for 8 KB I/O operations (Figure~\ref{fig:dio-lat-single-8kb}).
This latency gap fundamentally alters kernel I/O behavior, as the following stress conditions show.

\hspace{-10.0pt}{\bfseries Stress Condition 1: Limited Parallelism. }
Figure~\ref{fig:dio-test-threads} presents experimental results with
a fixed 8 KB I/O size and varying thread counts of 1, 2, 4, and 8.
The I/O size was set to 8 KB to align with PostgreSQL's typical I/O behavior, which operates in 8 KB blocks.

Under single-threaded access with small I/O sizes, distributed storage throughput drops to a tiny fraction of local storage performance.
Specifically, for random writes with a single thread, local SSD achieves 144 MB/s while EBS achieves 10.9 MB/s and Ceph RBD achieves approximately 3 MB/s.
This gap confirms that limited parallelism is a primary aggravating factor.

This symptom is directly relevant to database workloads: the kernel flusher thread operates as a single thread per device and therefore becomes latency-bound on high-overhead storage.

\hspace{-10.0pt}{\bfseries Stress Condition 2: Small I/O Size. }
Figure~\ref{fig:dio-test-iosize} shows results with a fixed thread count
of 1 and varying I/O sizes (8 KB, 32 KB, 128 KB, and 512 KB).
We use a single thread to isolate per-operation latency overheads,
which are especially relevant for workloads that rely on a small number
of kernel writeback threads.

Local SSDs maintain high bandwidth even at small I/O sizes; EBS and Ceph RBD, by contrast, suffer sharp throughput drops.
Even at 32 KB, EBS random write throughput remains below 50 MB/s, recovering to competitive levels only at 512 KB---demonstrating that distributed storage requires large I/O sizes to amortize per-operation latency.
Small I/O size is thus a second aggravating factor on distributed storage.

\hspace{-10.0pt}{\bfseries Stress Condition 3: Sustained Write Load with Buffered I/O. }
In direct I/O tests, fio results reflect both storage-side and client-side I/O performance.
However, buffered I/O writes data into memory first, making fio results insufficient for accurately measuring storage-side writeback performance.
Therefore, we utilized the iostat~\cite{noauthorIostat1Nodate} command with the extended
statistics option (-x) to measure storage-side I/O behavior under buffered
I/O, focusing on periods of valid activity and analyzing merge I/O statistics.
For the buffered I/O tests, the number of threads was set to 8, and the I/O size was fixed at 8 KB.

Table~\ref{tab:buf_fio} shows the results.
Under sustained random write workloads with buffered I/O, Ceph RBD throughput degrades to 51.1 MB/s,
which is substantially lower than the direct I/O result of 101 MB/s.
Closer inspection showed that dirty pages accumulated over time, eventually triggering I/O-less dirty throttling and degrading write throughput.
The kernel flusher thread cannot keep pace with sustained writes on high-latency storage.
The problem is further exacerbated by the KFT's periodic activation model---rather than running continuously, the KFT wakes at intervals determined by the kernel's dirty writeback parameters, creating gaps during which dirty pages accumulate unchecked.
On high-latency storage, each activation cycle flushes fewer pages due to per-I/O blocking, while dirty pages continue to grow between activations, accelerating the onset of I/O-less dirty throttling.

\hspace{-10.0pt}{\bfseries Stress Condition 4: Single-Volume QoS Ceiling. }
Distributed block storage is designed to provide stable I/O performance to multiple concurrent users through QoS mechanisms such as mClock.
Unlike local storage, where a single user can fully exploit the disk's capabilities, distributed storage enforces per-device performance guarantees that prevent any single volume from monopolizing the underlying physical infrastructure.
This design means that a single virtual disk cannot achieve the aggregate throughput available from the physical disk pool it spans.
When a database operates on a single volume, it is inherently constrained by this QoS ceiling regardless of the volume's nominal IOPS or bandwidth settings.
This condition interacts with Stress Conditions 1--3: even if the kernel flusher thread could process I/O faster, the single-volume QoS limit caps achievable throughput.

\subsection{Relief Conditions}
Equally important is identifying what relieves symptoms.

\hspace{-10.0pt}{\bfseries Relief Condition 1: Increased Parallelism. }
When thread count increases, distributed storage throughput improves dramatically.
As shown in Figure~\ref{fig:dio-test-threads}, EBS random write performance increases from 10.9 MB/s at 1 thread to 132 MB/s
at 4 threads---a 12x improvement.
Ceph RBD also shows substantial scaling, improving from approximately 3 MB/s to 101 MB/s---a 34x improvement.
At higher thread counts, the performance gap between distributed and local storage becomes much smaller.
Parallelization mitigates the pathology by amortizing per-I/O latency across concurrent operations.

\hspace{-10.0pt}{\bfseries Relief Condition 2: Larger I/O Size. }
Increasing I/O size from 8 KB to 512 KB significantly improves throughput
on distributed storage.
As shown in Figure~\ref{fig:dio-test-iosize},  EBS random write performance improves from 10.8 MB/s at 8 KB to 182 MB/s at 512 KB---a 17x improvement.
Ceph RBD shows similar improvement, from 2.8 MB/s to 67.7 MB/s---a 24x improvement.
I/O consolidation mitigates the pathology by reducing per-operation latency overhead.

\hspace{-10.0pt}{\bfseries Relief Condition 3: I/O Consolidation via Merge I/O. }

When merge I/O is effective, performance improves significantly.
Table~\ref{tab:buf_fio} shows that sequential write performance with buffered I/O can exceed direct I/O performance on distributed storage.
EBS achieves 487 MB/s with buffered I/O compared to 132 MB/s with direct I/O---a 3.7x improvement.
The average I/O size of 122 KB indicates that merge I/O successfully consolidates multiple 8 KB writes into larger operations.
Notably, EBS bandwidth exceeds its nominal IOPS-limited throughput through this consolidation---evidence that merge I/O is a potent mitigation mechanism.

\hspace{-10.0pt}{\bfseries Relief Condition 4: Multi-Volume Performance Aggregation. }
The preceding experiments already exhibit a per-volume ceiling: in Figure 2, EBS random write throughput saturates at 132 MB/s beyond 4 threads---the limit imposed by the configured 16,000 IOPS---while local SSD continues to scale.
This saturation is not caused by kernel-space bottlenecks but by a QoS constraint enforced per device.
Because storage costs are determined primarily by total volume size rather than volume count, allocating multiple standard gp3 volumes (each 3,000 IOPS, 125 MB/s) can aggregate to the same or higher throughput at equivalent cost, circumventing the single-volume ceiling.

These findings point to kernel-space I/O mechanisms and per-volume QoS constraints as the site of pathology.
In Section 4, we diagnose the specific root causes.

\section{DIAGNOSIS}\label{sec:proposed-method}
From the symptoms documented in Section 3, we diagnose four root causes that together create a pathological cascade.

\begin{figure*}[t]
  \centering
  \includegraphics[width=1.0\linewidth]{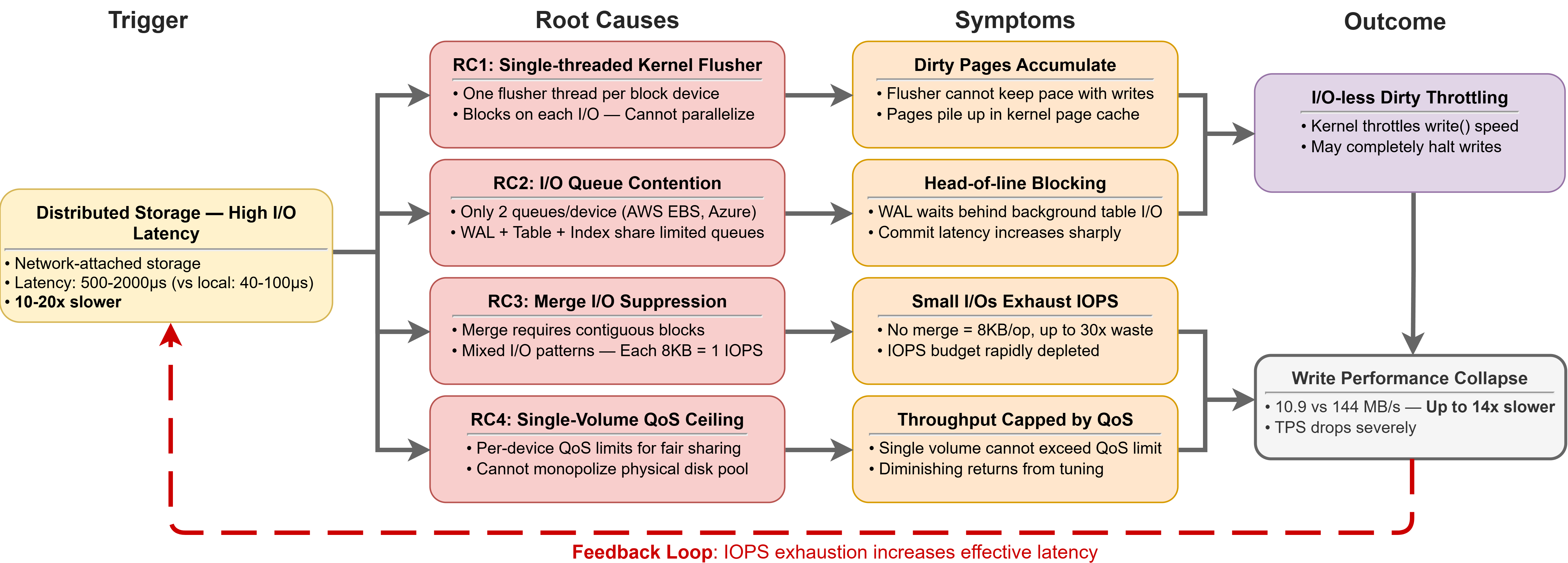}
  \vspace{-15pt}
  \caption{Pathological Cascade of Cloud Databases} 
  \vspace{-9pt}
  \label{fig:dialog}
\end{figure*}

\subsection{Root Causes}
{\bfseries Root Cause 1: Single-Threaded Kernel Flusher. }
The Linux kernel assigns one Kernel Flusher Thread (KFT) per block device. 
With local storage latency of 40--100 microseconds, this design works well. 
However, with distributed storage latency of 500--2000 microseconds, the single flusher thread becomes a severe bottleneck.

When database workloads execute numerous write system calls, data is written 
to memory as dirty pages.
The KFT flushes these pages, but as a single thread, it must wait for each write to complete before issuing the next.
With high-latency storage, dirty pages accumulate faster than they can be flushed, eventually triggering I/O-less Dirty Throttling that restricts or halts write system calls.
Databases such as PostgreSQL and MySQL issue non-WAL I/O in 8 KB or 16 KB pages, compounding this problem on distributed storage.

\hspace{-10.0pt}{\bfseries Root Cause 2: I/O Queue Contention. }
Cloud distributed block storage provides severely limited I/O queues---only two per device for both AWS EBS and Azure Disk Storage.
This constraint is a QoS design choice: cloud storage limits per-device queues to ensure stable performance for all tenants, rather than maximizing performance for any single instance.

When limited queues handle mixed I/O commands, latency-sensitive WAL writes compete with background table and index flushes.
Head-of-line blocking occurs when latency-sensitive WAL operations queue behind background table and index flushes, degrading transaction commit latency.

\hspace{-10.0pt}{\bfseries Root Cause 3: Merge I/O Suppression. }
Merge I/O consolidates adjacent requests into larger operations, as described in Section 2.3.5.
However, merge I/O requires contiguous blocks in the same I/O queue. In single-disk configurations, WAL, tables, and indexes interleave their I/O patterns, disrupting contiguity and suppressing merge I/O.
Without effective merging, most 8 KB writes consume one IOPS each, rapidly exhausting the IOPS budget.
This explains why increasing IOPS allocation often fails to resolve performance issues---the bottleneck is structural, not resource-limited.

\hspace{-10.0pt}{\bfseries Root Cause 4: Single-Volume QoS Ceiling. }
Distributed storage enforces per-device QoS guarantees to ensure fair resource sharing among tenants.
Unlike local storage, where one user can fully exploit the hardware, a single virtual disk cannot monopolize the physical disk pool it spans.
This architectural constraint imposes a throughput ceiling that is independent of the volume's configured IOPS or bandwidth. 
As observed in Section 3.2, even with 8 threads, Ceph RBD buffered I/O random write achieves only 51.1 MB/s---far below the cluster's aggregate performance (Table 1).
When a database operates on a single volume, it cannot leverage the broader physical infrastructure even when kernel-space bottlenecks (RC1--RC3) are partially mitigated.
This ceiling compounds the pathological cascade: even if flusher throughput were improved, a single volume's QoS limit would cap achievable I/O performance.

\vspace{-5pt}
\subsection{The Pathological Cascade}
These root causes form a reinforcing cycle: high latency slows the single-threaded flusher (RC1), causing dirty page accumulation.
Mixed I/O in limited queues (RC2) suppresses merge I/O (RC3), leaving most 8 KB writes to consume one IOPS each and further increasing effective latency.
The single-volume QoS ceiling (RC4) bounds maximum achievable throughput, so that tuning kernel parameters or increasing IOPS on a single disk provides limited improvement.
This cascade eventually triggers I/O-less dirty throttling, collapsing database write performance.
Previous research attributed cloud database performance challenges primarily to network PPS and bandwidth limitations.
The Aurora paper~\cite{verbitskiAmazon2017} designed storage-compute disaggregation to address these bottlenecks, and a subsequent reproduction study~\cite{pangUnderstanding2024} validated this perspective.
However, modern VM instances now offer 10--100 Gbps network bandwidth, and core networking equipment achieves 100--400 Gbps, making network far less likely to be the primary constraint.
Our analysis indicates that the pathology originates in kernel-space I/O mechanisms and distributed storage QoS constraints. 
Under this diagnosis, prior architectures appear to have circumvented these bottlenecks indirectly---transmitting only WAL from compute nodes to storage nodes---rather than addressing the underlying root causes.
In the next section, we describe SteelDB, an architecture that directly addresses these root causes.

\vspace{-7pt}
\section{TREATMENT: STEELDB}\label{subsec:steeldb}
This section describes SteelDB, a zero-patch architecture that directly addresses the root causes identified in Section 4.
We define a zero-patch architecture as one that improves database performance through cloud infrastructure configuration alone, without requiring any modifications to the kernel or database source code.

\vspace{-5pt}
\subsection{Architecture Overview}
\begin{figure}
  \centering
  \includegraphics[width=0.95\linewidth]{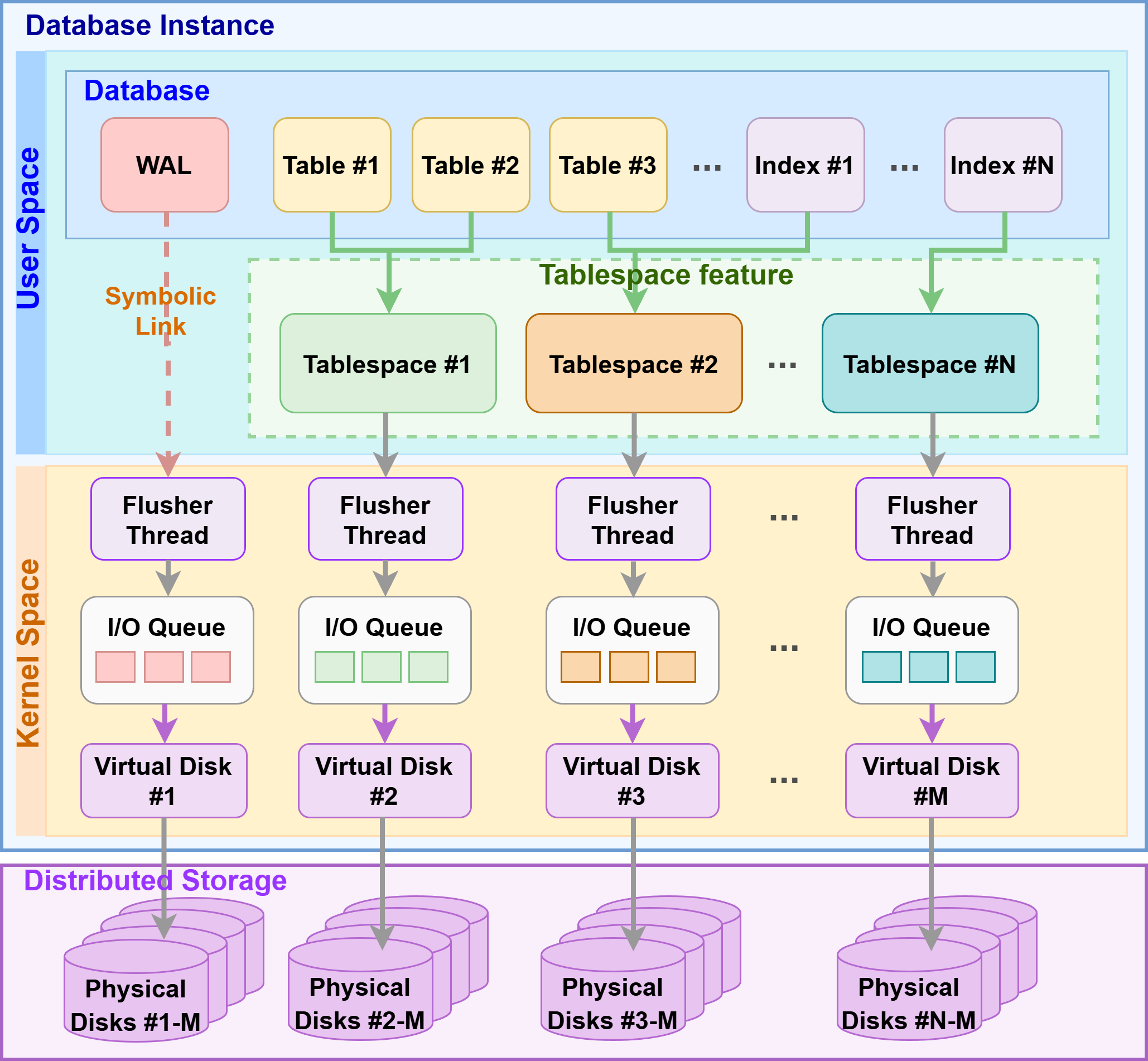}
    \vspace{-7pt}
  \caption{Architecture of SteelDB} 
  \vspace{-12pt}
  \label{fig:arch-steeldb}
\end{figure}

SteelDB allocates database components across multiple virtual disks.
Figure~\ref{fig:arch-steeldb} presents an overview of the architecture.
The key observation is that each virtual disk receives its own Kernel Flusher Thread, I/O queues, and data locality domain.
Distributing components across disks transforms these per-device constraints into parallel, isolated operations.
SteelDB utilizes standard PostgreSQL features---tablespaces for allocating tables and indexes to specific virtual disks, and symbolic links for placing the WAL directory on a dedicated disk. 
This approach requires no modifications to PostgreSQL or the Linux kernel, preserving full compatibility with upstream releases.

\subsection{Treatment Principles}
We derive four treatment principles from the diagnosis in Section 4.
All four can be realized using only standard database features---tablespaces and symbolic links---without modifying database or kernel code.

\hspace{-10.0pt}{\bfseries Principle 1: Parallelize Flusher Operations. }
Assigning WAL, tables, and indexes to separate virtual disks activates multiple Kernel Flusher Threads concurrently, converting the single-threaded flusher bottleneck into parallel writeback.
As demonstrated in Section 3, distributed storage achieves high throughput when accessed with multiple threads.
This allows the database to exploit the parallelism inherent in the distributed storage infrastructure, as demonstrated by Relief Condition 1.

\hspace{-10.0pt}{\bfseries Principle 2: Isolate I/O Queues by Function. }
Assigning each database component to its own virtual disk grants it dedicated I/O queues, eliminating the mixed-workload congestion diagnosed in RC2.
WAL writes, in particular, no longer queue behind background table and index flushes. 
Conversely, table and index read operations are no longer blocked by WAL fsync waits.

\hspace{-10.0pt}{\bfseries Principle 3: Restore Merge I/O Probability. }
Isolating database components onto dedicated volumes enables each device's I/O queues to process contiguous block ranges, maximizing merge I/O opportunities.
As demonstrated in Section 3, merging coalesces multiple 8 KB writes into larger operations, reducing IOPS consumption by an order of magnitude---delivering higher throughput at lower cost on standard cloud volumes.

\hspace{-10.0pt}{\bfseries Principle 4: Exploit Multi-Volume Performance Aggregation. }
Even if a future kernel patch were to resolve RC1 (single-threaded KFT), the cascade would still terminate at RC4 (QoS ceiling), confirming that distributed storage performance can only be truly unlocked through multi-volume horizontal scaling at the device layer.

Using multiple virtual disks, SteelDB exploits the per-device QoS guarantees of distributed storage.
Since each volume receives its own performance allocation, the aggregate throughput scales with the number of volumes.
This principle exploits the design characteristics of distributed storage: because storage costs are determined primarily by total volume size rather than volume count, multi-volume configurations achieve higher performance without additional cost.
Furthermore, multiple volumes span a broader set of physical disks across storage nodes, accessing infrastructure capacity that a single volume cannot reach.

\hspace{-10.0pt}{\bfseries Emergent Effect: Dirty Page Accumulation Prevention. }
The four principles reinforce one another.
Together, Principles 1---3 resolve the kernel-space bottlenecks while Principle 4 overcomes the infrastructure-level QoS constraint.
Together, they prevent dirty page accumulation---the trigger for I/O-less dirty throttling and write performance collapse.

\vspace{-5pt}
\subsection{Data Placement Strategy}
This section defines SteelDB's data placement algorithm. 
Like physical database design problems such as index selection and table partitioning~\cite{mozaffariSelfTuning2024}, optimal placement across virtual disks is inherently workload-dependent with no universal solution. 
Since most of the performance gain derives from the architectural separation itself, not from specific table assignments, we employ a practical scoring-based heuristic.

\hspace{-10.0pt}{\bfseries Step 1: Determine the Number of Flusher Threads. }
The initial number of KFTs is set to 4.
If the vCPU core count is 4 or fewer, it is set to 2.
The maximum number of KFTs is limited to one-fourth of the vCPU core count.
This constraint prevents performance degradation from excessive threads, which increase context-switching overhead and thread management costs.
In practice, 4 KFTs were sufficient across all configurations evaluated in this paper.

\hspace{-10.0pt}{\bfseries Step 2: Dedicated Resources for WAL. }
A dedicated I/O queue and KFT are assigned to WAL with the highest priority.
This allocation maximizes merge I/O efficiency for WAL writes and minimizes WAL I/O latency---both of which directly affect transaction commit time.
Additionally, a single KFT is initially allocated for all tables and indexes.

\hspace{-10.0pt}{\bfseries Step 3: Placement of Tables and Indexes. }
If the number of KFTs is 4 or more, a greedy scoring-based algorithm determines the allocation.
Let $t_i$ denote the $i$-th table or index.
A benchmark similar to the expected workload is executed to obtain the write frequency $W_{t_i}$, read frequency $R_{t_i}$, and data size $S_{t_i}$ of each $t_i$ using database statistical tools~\cite{noauthorPgbadgerNodate, noauthorPgStatsinfoNodate} and iostat.
The following score is then computed:
\begin{equation}
  Score(t_i) = \alpha \cdot N(W_{t_i}) + \beta \cdot N(R_{t_i}) + \gamma \cdot N(S_{t_i}),
\end{equation}
where $N(\cdot)$ normalizes values to $[0, 1]$, and weights are $\alpha = 0.5$, $\beta = 0.3$, and $\gamma = 0.2$.
Tables and indexes are sorted by score and assigned to KFTs in priority order.
To ensure balanced distribution, the Coefficient of Variation (CV) of scores across KFTs is checked: if CV exceeds 0.3 and the maximum KFT count has not been reached, an additional KFT is allocated and assignments are recalculated until the condition is met.

\hspace{-10.0pt}{\bfseries Step 4: Virtual Disk Performance Tuning. }
In cloud environments, I/O performance can be allocated to each volume based on workload characteristics.
High-performance I/O is expensive, making selective allocation essential. 
Ultra-low-latency and high-cost storage (e.g., io2) is ideal for WAL, as it reduces write latency and enhances transaction throughput.
In contrast, tables and indexes have lower latency requirements, making throughput and IOPS efficiency the primary concerns.
Using multiple cost-effective basic-tier virtual disks (e.g., gp3) instead of a single high-performance disk enables I/O optimization that balances performance and cost.

While this allocation problem resembles a multidimensional knapsack problem, optimal placement is less critical than realizing the SteelDB architecture and its treatment principles. 
A manually derived approximate solution using the proposed scoring algorithm provides sufficient benefits for practical deployments.
Dynamic rebalancing and online data migration under fluctuating workloads remain as promising directions for future work.

\vspace{-3pt}
\subsection{Leveraging Distributed Block Storage Characteristics}
SteelDB exploits distributed storage characteristics to exceed single-volume performance.

\hspace{-10.0pt}{\bfseries Multi-Volume Parallelism. }
As described in Principle 4 (Section 5.2), allocating multiple volumes enables aggregate performance that scales with volume count.
Distributed storage provides per-device performance guarantees that accumulate when multiple volumes are allocated to a single instance---yielding aggregate throughput beyond what any single volume can deliver.
For instance, using four 100GB virtual disks, each configured with the standard 3,000 IOPS and 125 MB/s bandwidth, achieves a combined performance of 12,000 IOPS and 500 MB/s.
The effectiveness of this strategy can be understood through distributed storage internals.
In Ceph, when a Placement Group can occupy a large amount of disk space, a virtual volume can span more physical disks than the replica count.
Hypothetically, if a single volume can access eight physical disks simultaneously, attaching 10 virtual disks could provide access to 80 physical disks.

\hspace{-10.0pt}{\bfseries Modern Cloud Infrastructure. }
Modern cloud environments support this strategy with storage-dedicated network bandwidth of 10--100 Gbps~\cite{awsEbsOptimized} and the ability to attach 27--128 virtual disks per instance~\cite{awsEbsVolumeLimits}.
Since storage costs are determined primarily by total volume size rather than the number of volumes, distributing 
data across multiple smaller disks incurs no additional cost.

\vspace{-3pt}
\subsection{Zero-Patch Advantages}
SteelDB's zero-patch architecture provides several practical advantages:

\hspace{-10.0pt}{\bfseries Diagnostic Clarity and Reproducibility. }
Because performance improvements derive solely from resource allocation rather than code modifications, they validate our diagnosis that kernel-space I/O mechanisms are the source of pathology.
Any PostgreSQL deployment can reproduce these results using standard features and publicly available cloud infrastructure.

\hspace{-10.0pt}{\bfseries Production Safety. }
Organizations requiring strict software integrity, such as government agencies and financial institutions, can 
adopt SteelDB without compromising security compliance.

\hspace{-10.0pt}{\bfseries Maintenance Elimination. }
Storage-compute disaggregated databases such as Aurora require substantial engineering effort to maintain 
compatibility with upstream PostgreSQL releases.
SteelDB eliminates this burden, enabling immediate adoption of new PostgreSQL versions and security patches.

\hspace{-10.0pt}{\bfseries Full Ecosystem Compatibility. }
SteelDB maintains complete compatibility with PostgreSQL tools, extensions, and monitoring solutions, preserving the full benefits of the open-source ecosystem.

\hspace{-10.0pt}{\bfseries Read Scalability. }
SteelDB inherits PostgreSQL's native read scaling through streaming replication.
Read replicas can be deployed without any modification, and because read scaling and SteelDB's write-path optimization operate independently, they together address both dimensions of OLTP scalability.

\section{EVALUATION}
We evaluate SteelDB through TPC-C benchmarks and comparisons with commercial cloud databases.

\subsection{Experimental Setup}\label{subsubsec:exp-setup-steeldb}
We use HammerDB~\cite{noauthorHammerdbNodate} with the TPC-C~\cite{noauthorTpcCNodate} workload, a widely adopted OLTP benchmark.
The database size was configured as WH1,000, and the number of virtual users was set to 100.
The benchmark execution time was 20 minutes, preceded by a 5-minute ramp-up period.
The evaluation metrics used in this study is NOPM (New Orders Per Minute).
We exclude MySQL because its benchmark results are subject to license restrictions~\cite{noauthorOtnNodate} and InnoDB's adaptive flushing conflicts with SteelDB's multi-disk strategy.
For SteelDB, the standard AWS EBS (gp3, 3,000 IOPS, 125 MB/s) was used. As baselines, we included a single-disk 
configuration, RAID 0 with four disks, a gp3 configuration with 12,000 IOPS and 500 MB/s, and a mixed two-disk configuration. 
The c6in.8xlarge VM instance (32vCPUs, 128 GB memory, 25 Gbps EBS bandwidth) was used for the database server.
The client instance was m6i.2xlarge (8vCPUs, 32 GB memory).
PostgreSQL configuration parameters are shown in Figure 8.
All kernel parameters, including dirty writeback settings, were left at their default values.
For Ceph RBD experiments, both database server and client were configured with 32vCPUs and 64 GB memory.
Figure~\ref{fig:ceph_servers} provides the Ceph cluster specifications.

\hspace{-10.0pt}{\bfseries Disk Allocation in SteelDB. }
For EBS, WAL was allocated to a dedicated disk with highest priority.
In the four-disk configuration, the stock table and order\_line table were allocated to separate disks based on 
their high write frequency.
For Ceph RBD, WAL was stored on SATA-connected SSDs while table and index data were stored on 
RBD.
All virtual disks were standardized to 200 GB volume size to ensure consistent performance.

\vspace{-3pt}
\subsection{Treatment Effectiveness on AWS EBS}
Table 2 presents the experimental results on AWS EBS.
We structure the analysis around the four root causes identified in Section 4.

\hspace{-10.0pt}{\bfseries Overall Performance. }
The four-disk SteelDB configuration (P2) achieved 242,107 NOPM, compared to 69,894 NOPM for the single-disk baseline (B1)---a 3.5x improvement.
The two-disk SteelDB configuration (P1) achieved 154,150 NOPM with only 6,000 IOPS and 250 MB/s, showing that even partial treatment yields measurable gains.

\hspace{-10.0pt}{\bfseries RC1 Validation: Single-Threaded Flusher. }
The four-disk RAID 0 configuration (B2) uses the same number of disks and aggregate specifications as SteelDB (P2), but employs only a single KFT due to RAID presenting one logical device to the kernel. RAID 0 achieved 123,981 NOPM versus SteelDB's 242,107 NOPM---a 2x gap.
This confirms that flusher parallelization (Principle 1), not added bandwidth alone, drives the improvement.
Table 4 further supports this finding: SteelDB showed higher CPU user utilization (14.8\% vs 10.4\%), indicating that the database processes spend more time on productive computation rather than waiting for I/O.

\hspace{-10.0pt}{\bfseries RC2 Validation: I/O Queue Contention. }
Table 4 shows that SteelDB's WAL disk average I/O queue size is 1.8, compared to 14.0 for the single enhanced disk configuration (B3).
This 8× reduction confirms that I/O queue isolation (Principle 2) eliminates the head-of-line blocking diagnosed in RC2.
With a dedicated queue, WAL writes complete with minimal queuing delay, directly improving transaction commit latency.

\hspace{-10.0pt}{\bfseries RC3 Validation: Merge I/O Suppression. }
In SteelDB's dedicated WAL disk, 56.2\% of I/O operations were merged, achieving an average I/O size of 75.5 KB---effectively consolidating approximately ten 8 KB WAL pages per operation.
This validates Principle 3: function-based data placement restores merge I/O probability.
In contrast, the single-disk configurations mix WAL, table, and index I/O patterns on shared queues, disrupting the contiguity required for merge I/O.

\hspace{-10.0pt}{\bfseries RC4 Validation: Single-Volume QoS Ceiling. }
SteelDB's four standard gp3 disks (P2: total 12,000 IOPS, 500 MB/s) outperformed a single enhanced gp3 disk with identical aggregate specifications (B3: 12,000 IOPS, 500 MB/s), achieving 242,107 versus 163,698 NOPM---a 1.5x gap.
The 1.5× gap between P2 and B3---despite identical aggregate IOPS and bandwidth---reflects contributions from all four root causes.
However, B3's inability to match P2 even with 12,000 IOPS on a single volume is consistent with the QoS ceiling diagnosed in RC4.

\hspace{-10.0pt}{\bfseries Synergistic Effect. }
The comparison between B4 (two-disk WAL split, 179,092 NOPM) and P2 (four-disk SteelDB, 242,107 NOPM) shows that the four principles combined yield greater improvement than addressing any root cause in isolation.
B4 separates WAL onto a dedicated disk, applying Principles 1 and 2 for WAL, but does not isolate table and index I/O.
SteelDB extends I/O isolation and merge I/O restoration to all database components, yielding 35\% additional improvement over partial treatment.

\vspace{-9pt}
\subsection{Treatment Effectiveness on Ceph RBD}
Table~\ref{tab:tpcc-ceph} Table 3 presents the results on Ceph RBD, which confirm the same pattern.
SteelDB RBDx4 achieved 384,062 NOPM, compared to 181,319 NOPM for the single RBD baseline---a 2.1x improvement.
In contrast, RAID 0 RBDx4 achieved only 300,639 NOPM despite using four disks.
The difference between SteelDB and RAID 0 confirms that I/O queue separation and flusher parallelization, not bandwidth aggregation alone, drive the improvement.
That these results hold across two independent storage implementations---EBS and Ceph RBD---supports the generality of our diagnosis: the root causes are inherent to the interaction between kernel I/O mechanisms and distributed block storage, not to any particular implementation.

\begin{table}[t]
  \centering
  \caption{TPC-C Benchmark Results on EBS}
  \vspace{-11pt}
  \label{tab:tpc-ebs}
  \resizebox{\linewidth}{!}{ %
  \begin{tabular}{p{2.75cm}cccc} %
    \toprule
    Method & KFTs & IOPS & Bandwidth(MB/s) & NOPM \\
    \midrule
    \textbf{B1} 1disk & 1 & 3,000 & 125 & 69,894 \\
    \textbf{B2} 4disks (RAID0) & 1 & 12,000 & 500 & 123,981 \\
    \textbf{B3} 1disk (Enhanced) & 1 & 12,000 & 500 & 163,698 \\
    \textbf{B4} 2disks (WAL Split) & 2 & 12,000 & 500 & 179,092 \\
    \textbf{P1} 2disks (SteelDB) & 2 & 6,000 & 250 & 154,150 \\
    \textbf{P2} 4disks (SteelDB) & 4 & 12,000 & 500 & 242,107 \\
    \bottomrule
  \end{tabular}
  } %
      \vspace{-9pt}
\end{table}

\begin{table}[t]
  \centering
  \caption{TPC-C Benchmark Results on Ceph RBD}
  \vspace{-11pt}
  \label{tab:tpcc-ceph}
  \resizebox{\linewidth}{!}{ %
  \begin{tabular}{p{5cm}ccc} %
    \toprule
    Method & WAL Disk & KFTs & NOPM\\
    \midrule
    \textbf{B1} SATA SSD & SATA SSD & 1 & 220,109 \\
    \textbf{B2} Ceph RBD x1 & SATA SSD & 2 & 181,319 \\
    \textbf{B3} Ceph RBD x4 (RAID0) & SATA SSD & 2 & 300,639 \\
    \textbf{P1} Ceph RBD x2 (SteelDB) & SATA SSD & 3 & 250,259 \\
    \textbf{P2} Ceph RBD x4 (SteelDB) & SATA SSD & 5 & 384,062 \\
    \bottomrule
  \end{tabular}
  } %
      \vspace{-9pt}
\end{table}

\begin{table}[t]
  \centering
  \caption{CPU Usage Analysis in Benchmarking on EBS}
  \vspace{-11pt}
  \label{tab:cpu_usuage_ana}
  \resizebox{\linewidth}{!}{ %
  \begin{tabular}{p{2.75cm}ccc} %
    \toprule
    \large{Method} & \large{\%usr} & \large{\%iowait} & \makecell{Average I/O Queue\\ Size in WAL Disk} \\
    \midrule
    \textbf{B3} 1disk (Enhanced) & 10.4 & 8.82 & 14.0 \\
    \textbf{P2} 4disks (SteelDB) & 14.8 & 20.1 & 1.8 \\
    \bottomrule
  \end{tabular}
  } %
      \vspace{-5pt}
\end{table}
\begin{figure}[t]
    \vspace{-2pt}
    \centering
    \fbox{%
        \begin{minipage}{8cm}
        \scriptsize
        \raggedright
        \texttt{shared\_buffers=4096MB, work\_mem=32MB, maintenance\_work\_mem=1024MB, checkpoint\_timeout=5min, checkpoint\_completion\_target=0.9, wal\_init\_zero=no, wal\_recycle=no, max\_wal\_size=8GB, min\_wal\_size=8GB}
        \end{minipage}%
    }
    \vspace{-10pt}
    \caption{PostgreSQL Configuration Parameters}
    \vspace{-13pt}
    \label{fig:postgresql_config}
\end{figure}

\vspace{-5pt}
\subsection{Comparison with Cloud Databases}
We compare SteelDB against commercially deployed storage-compute disaggregated databases.

\hspace{-10.0pt}{\bfseries Comparison with Amazon Aurora. }
We conduct a comparative evaluation of SteelDB against Amazon Aurora for PostgreSQL~\cite{noauthorOltpNodate},
a representative storage-compute disaggregated database.
Aurora was configured with I/O-Optimized~\cite{noauthorAmazonNodate2} storage,
which is officially designed for high I/O workloads and billed solely based on usage time regardless of I/O volume.
Since TPC-C is I/O-intensive with frequent small writes and commits, I/O-Optimized is the most favorable Aurora configuration for this workload---ensuring a fair comparison.

We evaluated two SteelDB configurations: a gp3-only setup (23 volumes, each 3,000 IOPS / 125 MB/s) and a hybrid setup (6 io2 volumes at 1,000 IOPS each + 17 gp3 volumes).
Volumes were grouped into four RAID 0 arrays via mdadm: 6 for WAL, 8 for stock table, 6 for order\_line table, and 3 for remaining tables, with chunk sizes of 8 KB for WAL and 32 KB for others.
The max\_wal\_size parameter was increased to 32 GB to align with improved disk performance.

The monthly cost calculations for SteelDB, Aurora, and AlloyDB are based on the pricing for the Tokyo region as of November 2024~\cite{noauthorAwsCalc}~\cite{noauthorGoogleCalc}.

\begin{table*}[t]
    \vspace{-7pt}
    \centering
    \caption{Performance Comparison of TPC-C Benchmark Results: SteelDB vs Other Cloud Databases}
    \vspace{-9pt}
    \label{tab:db_comparison}
    \renewcommand{\arraystretch}{1.05} %
    \resizebox{\textwidth}{!}{ %
    {\huge
    \begin{tabular}{l l l l r r r} %
        \toprule
        Databases & Instances & Disks & EBS Bandwidth & KFTs & NOPM & Monthly Cost (USD) \\
        \midrule
        Aurora  & r6i.4xlarge (16vCPU, 128GB) & I/O-Optimized  & Max 10Gbps & N/A & 236,011 & 2,657.2  \\
        SteelDB & r6i.4xlarge (16vCPU, 128GB) & gp3 x23        & Max 10Gbps  & 4 & 505,601  & 887.7   \\
        SteelDB & r6i.4xlarge (16vCPU, 128GB) & io2 x6 + gp3 x17 & Max 10Gbps & 4 & 580,256 & 1,331.7  \\
        \cmidrule(lr){1-7}
        Aurora  & r6i.8xlarge (32vCPU, 256GB) & I/O-Optimized  & 10Gbps    & N/A & 302,838 & 5,314.4  \\
        SteelDB & r6i.8xlarge (32vCPU, 256GB) & gp3 x23        & 10Gbps    & 4 & 637,532 & 1,775.4  \\
        SteelDB & r6i.8xlarge (32vCPU, 256GB) & io2 x6 + gp3 x17 & 10Gbps  & 4 & 926,589 & 2,219.4  \\
        SteelDB & c6in.8xlarge (32vCPU, 64GB)  & gp3 x23       & 25Gbps    & 4 & 638,790 & 1,667.9  \\
        SteelDB & c6in.8xlarge (32vCPU, 64GB)  & io2 x6 + gp3 x17 & 25Gbps & 4 & 930,640 & 2,111.9  \\
        \cmidrule(lr){1-7}
        AlloyDB & GCP Instance (64vCPU, 512GB) & Immutable & N/A      & N/A & 841,035  & 9,311.3  \\
        SteelDB & r6i.16xlarge (64vCPU, 512GB) & gp3 x23       & 20Gbps   & 4 & 640,556 & 3,550.7  \\
        SteelDB & r6i.16xlarge (64vCPU, 512GB) & io2 x6 + gp3 x17 & 20Gbps  & 4 & 1,210,515 & 3,994.7  \\
        \bottomrule
    \end{tabular}
    \vspace{-8pt}
    }}        \vspace{-8pt}
\end{table*}

Table 5 summarizes the results.
SteelDB outperformed Aurora across all configurations.
With the hybrid io2 configuration on r6i.8xlarge, SteelDB achieved 3.1x higher performance while reducing costs by 58\%---a 7.3× improvement in cost-performance ratio.
Even with only standard gp3 volumes, SteelDB on a 16vCPU instance outperformed Aurora on a 32vCPU instance by 67\%---achieving higher performance at approximately one-sixth the cost.
Comparing the c6in.8xlarge gp3 configuration against the single-disk baseline, SteelDB achieves a 9x performance improvement at no additional IOPS or bandwidth cost.

Notably, even in the configuration where SteelDB achieved 3.1x higher performance than Aurora, the network was not the primary bottleneck.
This experimentally confirms that the network bandwidth and PPS limitations identified as bottlenecks in prior research~\cite{verbitskiAmazon2017,pangUnderstanding2024} no longer constrain performance under modern cloud infrastructure---validating our diagnosis that the true bottleneck lies in kernel-space I/O mechanisms.

Furthermore, the TPC-C database size (WH1000) is approximately 100 GB.
On the c6in.8xlarge instance with only 64 GB memory---where the database does not fit in memory---SteelDB still outperformed Aurora, demonstrating that SteelDB's performance gains derive from I/O optimization rather than memory cache effects.

\hspace{-10.0pt}{\bfseries Comparison with GCP AlloyDB. }
In 2024, Google Cloud and Accenture published an official white paper~\cite{accentureAlloydb2024} evaluating AlloyDB against competing cloud databases including Amazon Aurora, OCI ATP~\cite{noauthorOracleNodate}, and OCI MySQL.
We adopt this as a reference because: (1) it is an official benchmark published by AlloyDB's own developer, ensuring optimally tuned configurations that favor AlloyDB; (2) it uses the same benchmarking methodology as our evaluation---HammerDB with the TPC-C model and WH1000 database size; and (3) AlloyDB outperformed all competing databases in their evaluation, establishing it as the strongest available baseline.
According to the white paper, AlloyDB was evaluated with 1, 16, 32, 64, 128, and 256 virtual users on a 64vCPU, 512 GB instance, achieving a peak performance of 841,035 NOPM at 256 users.
Performance declined at lower user counts.
In comparison, SteelDB recorded 1,210,515 NOPM with the same instance specification but only 100 virtual users. AlloyDB's own results show declining throughput at lower user counts, with its peak requiring 256 users.
SteelDB nonetheless achieved 1.4× higher throughput at 57\% lower cost.
SteelDB exceeded AlloyDB's peak even on half-sized instances: the r6i.8xlarge (32vCPU, 256 GB, 10 Gbps EBS bandwidth) achieved 926,589 NOPM, and the c6in.8xlarge (32vCPU, 64 GB, 25 Gbps EBS bandwidth) achieved 930,640 NOPM---both surpassing AlloyDB's 841,035 NOPM on a 64vCPU, 512 GB instance.
This confirms that the gain stems from I/O optimization, not hardware scaling.
Aurora and AlloyDB were evaluated under their most favorable conditions---I/O-Optimized storage and developer-tuned configurations, respectively---while all SteelDB runs use standard gp3 volumes at default settings (3,000 IOPS / 125 MB/s), the lowest-cost tier on AWS.
SteelDB outperformed both despite this asymmetry.
Further gains are achievable by scaling gp3 up to 16,000 IOPS (80,000 as of 2026~\cite{awsEbsVolumeTypes}), suggesting these results represent a lower bound of SteelDB's potential.

\begin{figure*}[t]
  \vspace{-0.4cm}
  \centering
  \includegraphics[width=0.7\textwidth]{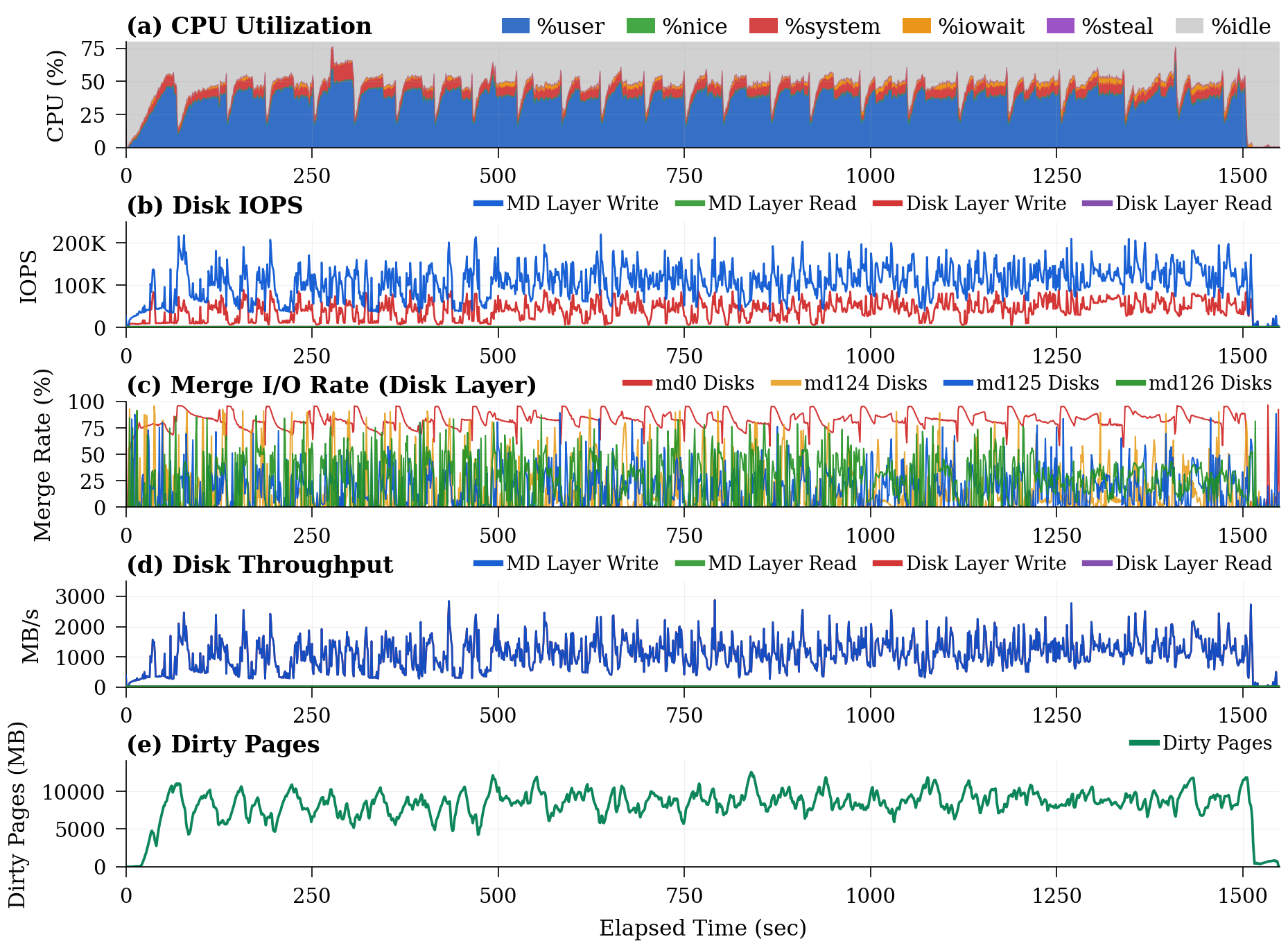}
    \vspace{-0.5cm}
  \caption{Resource Utilization During TPC-C Benchmark on SteelDB r6i.16xlarge (64vCPU, 512GB) with io2+gp3 Disks}
  \label{fig:resources}
  \vspace{-0.4cm}
\end{figure*}

\hspace{-10.0pt}{\bfseries Resource Utilization Analysis. }
Figure 9 presents resource utilization during the TPC-C benchmark on the r6i.16xlarge instance with the io2+gp3 hybrid disk configuration.
The disk I/O panels show almost exclusively write traffic, consistent with the fact that the WH1000 dataset (approximately 100 GB) resides entirely in the 512 GB main memory, eliminating storage-level read I/O. 
CPU utilization is dominated by \%usr; \%system occupies a non-trivial share due to KFT and I/O queue activity---a signature of SteelDB's multi-disk architecture.
The \%iowait remains low, consistent with the write-efficient multi-disk architecture and the dataset fitting entirely in memory.
Note that \%iowait on distributed storage should be interpreted with caution, as Linux does not attribute all network-induced wait time to this metric~\cite{Gregg2020}.
At the MD layer (RAID layer), aggregate I/O performance reached an average of 110,078 IOPS (peak 219,897) and 1,046 MB/s (peak 2,872 MB/s).
These values exceed nominal EBS performance limits, which is attributable to the efficient operation of merge I/O combined with the burst credit mechanism~\cite{park2020iops}---both within AWS specifications.
The actual disk-layer IOPS after merging average 42,366 (peak 89,929).
Merge I/O rates confirm that Principle 3 operates effectively: the WAL disk group achieves an average merge rate of 83.2\% (peak 96.0\%), while other disk groups range from 14.4\% to 25.3\% on average (peak 87.4\%--92.5\%).
Dirty page levels remain consistently low throughout the benchmark without any kernel parameter tuning, confirming that the four treatment principles work synergistically to prevent dirty page accumulation in production-scale workloads.

\hspace{-10.0pt}{\bfseries Applicability to Other Workloads. }
The four root causes diagnosed in Section 4 arise entirely within the kernel-storage interface and are agnostic to the workload running above it.
Any database that issues buffered I/O on distributed block storage is subject to the same pathological cascade, regardless of whether the access pattern is transactional, key-value, or analytical.
YCSB~\cite{Cooper2010}, for example, consists of simple key-value operations with minimal query processing overhead, making its throughput heavily dependent on raw storage I/O performance.
Write-heavy workloads would directly benefit from flusher parallelization and merge I/O restoration, while read-heavy workloads benefit from I/O queue isolation that eliminates head-of-line blocking.
When the dataset fits in memory, reads are served from the buffer cache---but this equally applies to Aurora and AlloyDB, so the performance differential remains determined by write-path efficiency, where SteelDB's optimization is most pronounced.
The aggregate I/O throughput in Figure 9 (average 1,046 MB/s, peak 2,872 MB/s) confirms that the treatment removes the structural I/O ceiling independent of access pattern.
Quantitative evaluation on additional benchmarks remains future work.

Taken together, these results establish that the decade-long trend toward storage-compute disaggregation was a response to kernel-space bottlenecks that can be resolved without proprietary storage or code modifications.
The true pathology lies in kernel-space I/O mechanisms, and SteelDB's zero-patch treatment delivers superior performance at a fraction of the cost---without the engineering burden of proprietary storage or patch maintenance.

\begin{table}[t]
\caption{Comparison of PostgreSQL Version Release Delays} %
\vspace{-12pt}
\label{tab:zero-patch-eval}
\centering
\small
\begin{tabular}{c|w{c}{2.5em}w{c}{2.5em}w{c}{2em}|w{c}{2.5em}w{c}{2.5em}w{c}{2em}}
\hline
\multirow{2}{*}{Database} & \multicolumn{3}{c|}{Major Version Delays [days]} & \multicolumn{3}{c}{Minor Version Delays [days]} \\
 & Avg & Max & Min & Avg & Max & Min \\ \hline
Aurora & 254 & 405 & 90 & 56 & 120 & 22  \\
AlloyDB & 426 & 438 & 405 & 153\small{*} & \small{N/A**} & \small{N/A**}  \\
SteelDB & $\approx$0 & $\approx$0 & $\approx$0 & $\approx$0 & $\approx$0 & $\approx$0  \\ \hline
\multicolumn{7}{l}{\small{*}\small{Based on AlloyDB 15.7 and 14.12 releases}} \\
\multicolumn{7}{l}{\small{**}\small{Cannot be calculated; many minor versions remain unreleased}} \\
\end{tabular}
\vspace{-11pt}
\end{table}

\subsection{Software Maintenance Cost Analysis}

The zero-patch architecture eliminates software maintenance overhead.
Storage-compute disaggregated databases must apply proprietary patches to each new PostgreSQL release.
We analyze the release schedules of Aurora~\cite{noauthorReleaseNodate} and AlloyDB~\cite{noauthorAlloydbNodate1} to quantify this burden.
Table 6 summarizes the results.
The PostgreSQL community releases major versions annually and minor versions quarterly.
As of March 3, 2025---over five months after the PostgreSQL 17 release on September 26, 2024---neither Aurora nor AlloyDB supports PostgreSQL 17, while SteelDB can deploy it immediately upon release.
Aurora's major version delays average 254 days (max 405), with minor version delays averaging 56 days.
AlloyDB exhibits even larger delays: major version delays average 426 days (max 438), exceeding the annual release cadence itself, and many minor versions remain unreleased entirely.
These delays persist despite the engineering resources of AWS and Google Cloud---the world's top-tier cloud providers---suggesting that patch-based maintenance carries inherently high cost.
These delays slow the response to security vulnerabilities---a concern for government agencies and financial institutions subject to strict compliance requirements.
SteelDB eliminates these costs, enabling immediate deployment of new PostgreSQL versions with binary-level identity to upstream releases.

\vspace{-3pt}
\subsection{Discussion on Availability}
A natural concern is that increasing the number of virtual disks introduces additional points of failure, thereby reducing system durability. We prove mathematically that this intuition is incorrect: volume partitioning has no effect on system-wide failure probability.
Virtual disks in distributed storage are managed through software-based quotas rather than physical disk boundaries.
Let block-level failure probability be $p$, replica count $r$, and total capacity $B$ blocks divided into $k$ volumes. Each volume's failure probability is $P_i = 1 - (1 - p^r)^{B/k}$.
The system-wide failure probability is:

\vspace{-6pt}
\begin{equation}
P_{\text{system}} = 1 - (1 - P_i)^k = 1 - (1 - p^r)^B \,,
\end{equation}

\noindent The $k$ terms cancel, proving that $P_{\text{system}}$ depends only on total capacity $B$, not on partition count $k$.
Furthermore, smaller volumes reduce the blast radius of individual failures and shorten recovery time.
In practice, distributed storage achieves high durability.
AWS gp3 volumes report an Annual Failure Rate (AFR) below 0.2\%~\cite{brooker2020millions}, and io2 volumes below 0.001\%~\cite{awsEbsVolumeTypes}---significantly lower than the typical 1\% AFR of standard HDDs.

For systems requiring additional availability, standard measures such as streaming replication across availability zones, WAL archiving to external storage (e.g., S3), or reducing virtual disk count by purchasing additional IOPS can be applied.
These measures are orthogonal to SteelDB and provide availability comparable to traditional configurations.

Regarding operational complexity, provisioning and mounting multiple volumes is a one-time operation easily automated with tools such as Terraform~\cite{hashicorpTerraform}, Pulumi~\cite{pulumiPulumi}, and Ansible~\cite{redhatAnsible}.
If SteelDB were offered as a managed service, users would not see this complexity at all---just as Aurora users are unaware of its internal multi-node storage architecture.

\section{RELATED WORK}
{\bfseries Storage-Compute Disaggregated Databases. }
Amazon Aurora~\cite{verbitskiAmazon2017}, released in 2015, established the template for subsequent cloud-native databases including Azure Hyperscale~\cite{antonopoulosSocrates2019}, PolarDB~\cite{caoPolardb2021}, Neon~\cite{noauthorNeonNodate}, and GCP AlloyDB~\cite{noauthorAlloydbNodate}.
Aurora introduced an architecture that separates compute and storage layers, writing only WAL from compute nodes to storage nodes where data pages are reconstructed.
The Aurora paper attributes the design's write performance gains to avoiding network bandwidth and PPS bottlenecks.
Several aspects of Aurora's effectiveness remained open to question, prompting a reproduction study at SIGMOD 2024~\cite{pangUnderstanding2024} that independently validated key design choices and clarified its performance behavior.
However, neither the Aurora paper nor the reproduction study addresses the kernel flusher thread bottleneck caused by cloud distributed storage latency at all, nor do they discuss I/O queue contention or merge I/O suppression within the kernel space at all---the root causes identified in this paper.
These kernel-space bottlenecks have been entirely overlooked in prior research.
Rather than identifying and addressing these root causes, prior architectures adopted WAL-only writes from compute nodes, indirectly circumventing the kernel-space pathology without recognizing its existence.
Our analysis shows that the degradation attributed to network limitations is more precisely explained by kernel-space I/O mechanisms.
SteelDB addresses these root causes directly, using general-purpose distributed block storage with a zero-patch architecture. 
This delivers both cost efficiency and rapid security updates---advantages that proprietary-storage approaches, burdened with patch maintenance, cannot match.

\hspace{-10.0pt}{\bfseries Distributed Transactional Databases. }
Distributed transactional databases such as CockroachDB~\cite{taftCockroachdb2020},
Spanner~\cite{corbettSpanner2013}, PolarDB-X~\cite{caoPolardbX2022}, TiDB~\cite{huangTidb2020}, YugabyteDB~\cite{noauthorDistributedNodate}, and OceanBase~\cite{yangOceanbase2022} achieve scalable
performance and high availability by distributing transaction consensus across multiple nodes.
However, these systems incur overhead on common SQL patterns---distributed JOINs, GROUP BY---limiting their use as general-purpose OLTP databases.
On cloud infrastructure, they face a dual latency problem: network-based consensus among nodes and network-based access to distributed block storage compound to create significantly higher per-operation latency, reducing per-node performance.
Our work addresses a different problem and takes a fundamentally different approach.

\hspace{-10.0pt}{\bfseries Shared-Storage Databases. }
Shared-storage databases such as Oracle RAC~\cite{chandrasekaranShared2003} and PolarDB-MP~\cite{yangPolardbMp2024} enable multi-primary access via shared storage but require complex distributed lock management and high-performance shared storage that incurs extremely high costs in cloud environments.

\hspace{-10.0pt}{\bfseries Database and Kernel I/O Evolution, and CloudJump. }
Before hardware RAID controllers with write-back cache became common in the early 1990s, databases distributed data across multiple disks via tablespaces for I/O parallelism.
With the introduction of hardware RAID controllers, the standard practice shifted toward single-volume RAID arrays, and this single-volume approach remains dominant today---Amazon RDS~\cite{noauthorManagedNodate}, Google Cloud SQL~\cite{noauthorCloudNodate1}, and Azure Database do not support multi-volume architectures.

Meanwhile, the Linux kernel has evolved significantly.
A 2009 patch~\cite{noauthorWritebackNodate} improved Kernel Flusher Thread management by assigning one KFT per disk,
whereas previously a single KFT was shared among all disks.
Multi-queue block I/O (blk-mq) support was introduced in a 2013 patch~\cite{noauthorBlkMqNodate} and became the
default in kernel 4.19~\cite{noauthorScsiNodate}, enabling parallel I/O processing.
Recent research has also addressed buffered I/O optimization~\cite{kimRevitalizing2023} and the combination of buffered and direct I/O in distributed file systems~\cite{qianCombining2024}.
SteelDB builds on these kernel improvements---per-device KFT and multi-queue I/O---while addressing the specific characteristics of cloud distributed storage. Strategic data placement atop these primitives yields performance gains that simple data distribution cannot achieve.
Thus, SteelDB is not a revival of legacy multi-disk placement for local bandwidth---it addresses kernel-space bottlenecks that exist only on distributed block storage, a configuration no cloud managed database supports today.

CloudJump~\cite{chenCloudjump2022} accelerates MySQL on cloud block storage via direct I/O patches, but does not address the KFT bottleneck or I/O queue limitations identified here.
\enlargethispage{2\baselineskip}

\section{CONCLUSION}
We presented a pathological analysis of write performance degradation in cloud OLTP databases.
From controlled experiments on distributed block storage, we identified four root causes—single-threaded flusher limitations, I/O queue contention, merge I/O suppression, and single-volume QoS ceiling.
Based on this diagnosis, we validated SteelDB, a cross-layer zero-patch orchestration architecture that addresses these root causes by coordinating database, kernel, and distributed block storage layers through strategic data placement alone.
The pathological cascade emerges from the interaction among these layers---invisible when any single layer is examined in isolation, which explains why prior work overlooked these bottlenecks.
TPC-C evaluations demonstrated that SteelDB achieves up to 9x performance improvement.
Compared to Aurora, SteelDB achieved 3.1x higher performance while reducing costs by 58\%, resulting in a 7.3x improvement in cost efficiency.
Against AlloyDB, it achieved 1.4x higher throughput at 57\% lower cost.
These results were achieved using the lowest-cost configuration of standard gp3 and io2 volumes.
The zero-patch architecture eliminates software maintenance overhead: Aurora averages 254 days for major version upgrades; SteelDB enables immediate adoption.
By exposing bottlenecks that arise from the interaction between buffered I/O and high-latency distributed storage, this work shows that high-performance cloud databases can be built on commodity kernel primitives---without proprietary storage or custom patches.
This perspective opens a new direction for cloud database research.
\vfill
\clearpage

{
\bibliographystyle{ACM-Reference-Format}
\bibliography{steeldb20250317}


\begin{thebibliography}{67}


\ifx \showCODEN    \undefined \def \showCODEN     #1{\unskip}     \fi
\ifx \showISBNx    \undefined \def \showISBNx     #1{\unskip}     \fi
\ifx \showISBNxiii \undefined \def \showISBNxiii  #1{\unskip}     \fi
\ifx \showISSN     \undefined \def \showISSN      #1{\unskip}     \fi
\ifx \showLCCN     \undefined \def \showLCCN      #1{\unskip}     \fi
\ifx \shownote     \undefined \def \shownote      #1{#1}          \fi
\ifx \showarticletitle \undefined \def \showarticletitle #1{#1}   \fi
\ifx \showURL      \undefined \def \showURL       {\relax}        \fi
\providecommand\bibfield[2]{#2}
\providecommand\bibinfo[2]{#2}
\providecommand\natexlab[1]{#1}
\providecommand\showeprint[2][]{arXiv:#2}

\bibitem[Aghayev et~al\mbox{.}(2019)]%
        {aghayevFile2019}
\bibfield{author}{\bibinfo{person}{Abutalib Aghayev}, \bibinfo{person}{Sage
  Weil}, \bibinfo{person}{Michael Kuchnik}, \bibinfo{person}{Mark Nelson},
  \bibinfo{person}{Gregory~R. Ganger}, {and} \bibinfo{person}{George
  Amvrosiadis}.} \bibinfo{year}{2019}\natexlab{}.
\newblock \showarticletitle{File systems unfit as distributed storage backends:
  lessons from 10 years of {Ceph} evolution}. In
  \bibinfo{booktitle}{\emph{Proceedings of the 27th {ACM} {Symposium} on
  {Operating} {Systems} {Principles}. (SOSP '19)}}. \bibinfo{publisher}{ACM},
  \bibinfo{address}{New York, NY, USA}, \bibinfo{pages}{353--369}.
\newblock


\bibitem[{Amazon Web Services}(2024a)]%
        {noauthorAmazonNodate2}
\bibfield{author}{\bibinfo{person}{{Amazon Web Services}}.}
  \bibinfo{year}{2024}\natexlab{a}.
\newblock \bibinfo{title}{Amazon {Aurora} storage - {Amazon} {Aurora}}.
\newblock
\urldef\tempurl%
\url{https://docs.aws.amazon.com/AmazonRDS/latest/AuroraUserGuide/Aurora.Overview.StorageReliability.html#aurora-storage-type}
\showURL{%
\tempurl}
\newblock
\shownote{Accessed: March 30, 2026}.


\bibitem[{Amazon Web Services}(2024b)]%
        {awsEbsOptimized}
\bibfield{author}{\bibinfo{person}{{Amazon Web Services}}.}
  \bibinfo{year}{2024}\natexlab{b}.
\newblock \bibinfo{title}{Amazon {EBS}-Optimized Instance Types}.
\newblock
\urldef\tempurl%
\url{https://docs.aws.amazon.com/AWSEC2/latest/UserGuide/ebs-optimized.html}
\showURL{%
\tempurl}
\newblock
\shownote{Accessed: March 30, 2026}.


\bibitem[{Amazon Web Services}(2024c)]%
        {awsEbsVolumeLimits}
\bibfield{author}{\bibinfo{person}{{Amazon Web Services}}.}
  \bibinfo{year}{2024}\natexlab{c}.
\newblock \bibinfo{title}{Amazon {EBS} Volume Limits for {Amazon EC2}
  Instances}.
\newblock
\urldef\tempurl%
\url{https://docs.aws.amazon.com/AWSEC2/latest/UserGuide/volume_limits.html}
\showURL{%
\tempurl}
\newblock
\shownote{Accessed: March 30, 2026}.


\bibitem[{Amazon Web Services}(2024d)]%
        {awsEbsVolumeTypes}
\bibfield{author}{\bibinfo{person}{{Amazon Web Services}}.}
  \bibinfo{year}{2024}\natexlab{d}.
\newblock \bibinfo{title}{Amazon EBS volume types - Amazon Elastic Compute
  Cloud}.
\newblock
\urldef\tempurl%
\url{https://docs.aws.amazon.com/AWSEC2/latest/UserGuide/ebs-volume-types.html}
\showURL{%
\tempurl}
\newblock
\shownote{Accessed: March 30, 2026}.


\bibitem[{Amazon Web Services}(2024e)]%
        {noauthorAwsCalc}
\bibfield{author}{\bibinfo{person}{{Amazon Web Services}}.}
  \bibinfo{year}{2024}\natexlab{e}.
\newblock \bibinfo{title}{{AWS} {Pricing} {Calculator}}.
\newblock
\urldef\tempurl%
\url{https://calculator.aws/#/}
\showURL{%
\tempurl}
\newblock
\shownote{Accessed: March 30, 2026}.


\bibitem[{Amazon Web Services}(2024f)]%
        {noauthorCloudNodate}
\bibfield{author}{\bibinfo{person}{{Amazon Web Services}}.}
  \bibinfo{year}{2024}\natexlab{f}.
\newblock \bibinfo{title}{Cloud {Block} {Storage} - {Amazon} {EBS} - {AWS}}.
\newblock
\urldef\tempurl%
\url{https://aws.amazon.com/ebs/}
\showURL{%
\tempurl}
\newblock
\shownote{Accessed: March 30, 2026}.


\bibitem[{Amazon Web Services}(2024g)]%
        {noauthorManagedNodate}
\bibfield{author}{\bibinfo{person}{{Amazon Web Services}}.}
  \bibinfo{year}{2024}\natexlab{g}.
\newblock \bibinfo{title}{Managed {SQL} {Database} - {Amazon} {Relational}
  {Database} {Service} ({RDS}) - {AWS}}.
\newblock
\urldef\tempurl%
\url{https://aws.amazon.com/rds/}
\showURL{%
\tempurl}
\newblock
\shownote{Accessed: March 30, 2026}.


\bibitem[{Amazon Web Services}(2024h)]%
        {noauthorOltpNodate}
\bibfield{author}{\bibinfo{person}{{Amazon Web Services}}.}
  \bibinfo{year}{2024}\natexlab{h}.
\newblock \bibinfo{title}{{OLTP} {Database}, {MySQL} {And} {PostgreSQL}
  {Managed} {Database} - {Amazon} {Aurora} - {AWS}}.
\newblock
\urldef\tempurl%
\url{https://aws.amazon.com/rds/aurora/?nc1=h_ls}
\showURL{%
\tempurl}
\newblock
\shownote{Accessed: March 30, 2026}.


\bibitem[{Amazon Web Services}(2024i)]%
        {noauthorReleaseNodate}
\bibfield{author}{\bibinfo{person}{{Amazon Web Services}}.}
  \bibinfo{year}{2024}\natexlab{i}.
\newblock \bibinfo{title}{Release calendars for {Aurora} {PostgreSQL} -
  {Amazon} {Aurora}}.
\newblock
\urldef\tempurl%
\url{https://docs.aws.amazon.com/AmazonRDS/latest/AuroraPostgreSQLReleaseNotes/aurorapostgresql-release-calendar.html}
\showURL{%
\tempurl}
\newblock
\shownote{Accessed: March 30, 2026}.


\bibitem[Antonopoulos et~al\mbox{.}(2019)]%
        {antonopoulosSocrates2019}
\bibfield{author}{\bibinfo{person}{Panagiotis Antonopoulos},
  \bibinfo{person}{Alex Budovski}, \bibinfo{person}{Cristian Diaconu},
  \bibinfo{person}{Alejandro Hernandez~Saenz}, \bibinfo{person}{Jack Hu},
  \bibinfo{person}{Hanuma Kodavalla}, \bibinfo{person}{Donald Kossmann},
  \bibinfo{person}{Sandeep Lingam}, \bibinfo{person}{Umar~Farooq Minhas},
  \bibinfo{person}{Naveen Prakash}, \bibinfo{person}{Vijendra Purohit},
  \bibinfo{person}{Hugh Qu}, \bibinfo{person}{Chaitanya~Sreenivas Ravella},
  \bibinfo{person}{Krystyna Reisteter}, \bibinfo{person}{Sheetal Shrotri},
  \bibinfo{person}{Dixin Tang}, {and} \bibinfo{person}{Vikram Wakade}.}
  \bibinfo{year}{2019}\natexlab{}.
\newblock \showarticletitle{Socrates: {The} {New} {SQL} {Server} in the
  {Cloud}}. In \bibinfo{booktitle}{\emph{Proceedings of the ACM International
  Conference on Management of Data. (SIGMOD '19)}}. \bibinfo{publisher}{ACM},
  \bibinfo{address}{New York, NY, USA}, \bibinfo{pages}{1743--1756}.
\newblock


\bibitem[Axboe(2013)]%
        {noauthorBlkMqNodate}
\bibfield{author}{\bibinfo{person}{Jens Axboe}.}
  \bibinfo{year}{2013}\natexlab{}.
\newblock \bibinfo{title}{blk-mq: new multi-queue block {IO} queueing mechanism
  \textperiodcentered{} torvalds/linux@320ae51}.
\newblock
\urldef\tempurl%
\url{https://github.com/torvalds/linux/commit/320ae51feed5c2f13664aa05a76bec198967e04d}
\showURL{%
\tempurl}
\newblock
\shownote{Accessed: March 30, 2026}.


\bibitem[Axboe(2017)]%
        {noauthorScsiNodate}
\bibfield{author}{\bibinfo{person}{Jens Axboe}.}
  \bibinfo{year}{2017}\natexlab{}.
\newblock \bibinfo{title}{scsi: default to scsi-mq \textperiodcentered{}
  torvalds/linux@5c279bd}.
\newblock
\urldef\tempurl%
\url{https://github.com/torvalds/linux/commit/5c279bd9e40624f4ab6e688671026d6005b066fa}
\showURL{%
\tempurl}
\newblock
\shownote{Accessed: March 30, 2026}.


\bibitem[Axboe(2024)]%
        {axboeFlexible2022}
\bibfield{author}{\bibinfo{person}{Jens Axboe}.}
  \bibinfo{year}{2024}\natexlab{}.
\newblock \bibinfo{title}{Flexible {I}/{O} {Tester}}.
\newblock
\urldef\tempurl%
\url{https://github.com/axboe/fio}
\showURL{%
\tempurl}
\newblock
\shownote{Accessed: March 30, 2026}.


\bibitem[Brooker et~al\mbox{.}(2020)]%
        {brooker2020millions}
\bibfield{author}{\bibinfo{person}{Marc Brooker}, \bibinfo{person}{Tao Chen},
  {and} \bibinfo{person}{Fan Ping}.} \bibinfo{year}{2020}\natexlab{}.
\newblock \showarticletitle{Millions of Tiny Databases}. In
  \bibinfo{booktitle}{\emph{17th USENIX Symposium on Networked Systems Design
  and Implementation (NSDI 20)}}. \bibinfo{publisher}{USENIX Association},
  \bibinfo{address}{Santa Clara, CA}, \bibinfo{pages}{463--478}.
\newblock


\bibitem[Cao et~al\mbox{.}(2022)]%
        {caoPolardbX2022}
\bibfield{author}{\bibinfo{person}{Wei Cao}, \bibinfo{person}{Feifei Li},
  \bibinfo{person}{Gui Huang}, \bibinfo{person}{Jianghang Lou},
  \bibinfo{person}{Jianwei Zhao}, \bibinfo{person}{Dengcheng He},
  \bibinfo{person}{Mengshi Sun}, \bibinfo{person}{Yingqiang Zhang},
  \bibinfo{person}{Sheng Wang}, \bibinfo{person}{Xueqiang Wu},
  \bibinfo{person}{Han Liao}, \bibinfo{person}{Zilin Chen},
  \bibinfo{person}{Xiaojian Fang}, \bibinfo{person}{Mo Chen},
  \bibinfo{person}{Chenghui Liang}, \bibinfo{person}{Yanxin Luo},
  \bibinfo{person}{Huanming Wang}, \bibinfo{person}{Songlei Wang},
  \bibinfo{person}{Zhanfeng Ma}, \bibinfo{person}{Xinjun Yang},
  \bibinfo{person}{Xiang Peng}, \bibinfo{person}{Yubin Ruan},
  \bibinfo{person}{Yuhui Wang}, \bibinfo{person}{Jie Zhou},
  \bibinfo{person}{Jianying Wang}, \bibinfo{person}{Qingda Hu}, {and}
  \bibinfo{person}{Junbin Kang}.} \bibinfo{year}{2022}\natexlab{}.
\newblock \showarticletitle{{PolarDB}-{X}: {An} {Elastic} {Distributed}
  {Relational} {Database} for {Cloud}-{Native} {Applications}}. In
  \bibinfo{booktitle}{\emph{{IEEE} {International} {Conference} on {Data}
  {Engineering}. (ICDE '22)}}. \bibinfo{publisher}{IEEE},
  \bibinfo{address}{Kuala Lumpur, Malaysia}, \bibinfo{pages}{2859--2872}.
\newblock


\bibitem[Cao et~al\mbox{.}(2021)]%
        {caoPolardb2021}
\bibfield{author}{\bibinfo{person}{Wei Cao}, \bibinfo{person}{Yingqiang Zhang},
  \bibinfo{person}{Xinjun Yang}, \bibinfo{person}{Feifei Li},
  \bibinfo{person}{Sheng Wang}, \bibinfo{person}{Qingda Hu},
  \bibinfo{person}{Xuntao Cheng}, \bibinfo{person}{Zongzhi Chen},
  \bibinfo{person}{Zhenjun Liu}, \bibinfo{person}{Jing Fang},
  \bibinfo{person}{Bo Wang}, \bibinfo{person}{Yuhui Wang},
  \bibinfo{person}{Haiqing Sun}, \bibinfo{person}{Ze Yang},
  \bibinfo{person}{Zhushi Cheng}, \bibinfo{person}{Sen Chen},
  \bibinfo{person}{Jian Wu}, \bibinfo{person}{Wei Hu}, \bibinfo{person}{Jianwei
  Zhao}, \bibinfo{person}{Yusong Gao}, \bibinfo{person}{Songlu Cai},
  \bibinfo{person}{Yunyang Zhang}, {and} \bibinfo{person}{Jiawang Tong}.}
  \bibinfo{year}{2021}\natexlab{}.
\newblock \showarticletitle{{PolarDB} {Serverless}: {A} {Cloud} {Native}
  {Database} for {Disaggregated} {Data} {Centers}}. In
  \bibinfo{booktitle}{\emph{Proceedings of the ACM International Conference on
  Management of Data. (SIGMOD '21)}}. \bibinfo{publisher}{ACM},
  \bibinfo{address}{New York, NY, USA}, \bibinfo{pages}{2477--2489}.
\newblock


\bibitem[Celko(2010)]%
        {celkoJoe2010}
\bibfield{author}{\bibinfo{person}{Joe Celko}.}
  \bibinfo{year}{2010}\natexlab{}.
\newblock \bibinfo{booktitle}{\emph{Joe {Celko}'s {SQL} for {Smarties},
  {Fourth} {Edition}: {Advanced} {SQL} {Programming}}}.
\newblock \bibinfo{publisher}{Morgan Kaufmann}, \bibinfo{address}{Burlington,
  MA}.
\newblock


\bibitem[Chandrasekaran and Bamford(2003)]%
        {chandrasekaranShared2003}
\bibfield{author}{\bibinfo{person}{S. Chandrasekaran} {and} \bibinfo{person}{R.
  Bamford}.} \bibinfo{year}{2003}\natexlab{}.
\newblock \showarticletitle{Shared cache - the future of parallel databases}.
  In \bibinfo{booktitle}{\emph{Proceedings 19th {International} {Conference} on
  {Data} {Engineering}. (ICDE '03)}}. \bibinfo{publisher}{IEEE},
  \bibinfo{address}{Bangalore, India}, \bibinfo{pages}{840--850}.
\newblock


\bibitem[Chen et~al\mbox{.}(2022)]%
        {chenCloudjump2022}
\bibfield{author}{\bibinfo{person}{Zongzhi Chen}, \bibinfo{person}{Xinjun
  Yang}, \bibinfo{person}{Feifei Li}, \bibinfo{person}{Xuntao Cheng},
  \bibinfo{person}{Qingda Hu}, \bibinfo{person}{Zheyu Miao},
  \bibinfo{person}{Rongbiao Xie}, \bibinfo{person}{Xiaofei Wu},
  \bibinfo{person}{Kang Wang}, \bibinfo{person}{Zhao Song},
  \bibinfo{person}{Haiqing Sun}, \bibinfo{person}{Zechao Zhuang},
  \bibinfo{person}{Yuming Yang}, \bibinfo{person}{Jie Xu},
  \bibinfo{person}{Liang Yin}, \bibinfo{person}{Wenchao Zhou}, {and}
  \bibinfo{person}{Sheng Wang}.} \bibinfo{year}{2022}\natexlab{}.
\newblock \showarticletitle{{CloudJump}: {Optimizing} {Cloud} {Databases} for
  {Cloud} {Storages}}.
\newblock \bibinfo{journal}{\emph{Proceedings of the VLDB Endowment. (PVLDB)}}
  \bibinfo{volume}{15}, \bibinfo{number}{12} (\bibinfo{year}{2022}),
  \bibinfo{pages}{3432--3444}.
\newblock


\bibitem[Cooper et~al\mbox{.}(2010)]%
        {Cooper2010}
\bibfield{author}{\bibinfo{person}{Brian~F. Cooper}, \bibinfo{person}{Adam
  Silberstein}, \bibinfo{person}{Erwin Tam}, \bibinfo{person}{Raghu
  Ramakrishnan}, {and} \bibinfo{person}{Russell Sears}.}
  \bibinfo{year}{2010}\natexlab{}.
\newblock \showarticletitle{Benchmarking Cloud Serving Systems with {YCSB}}. In
  \bibinfo{booktitle}{\emph{Proceedings of the 1st ACM Symposium on Cloud
  Computing}} \emph{(\bibinfo{series}{SoCC '10})}. \bibinfo{publisher}{ACM},
  \bibinfo{address}{New York, NY, USA}, \bibinfo{pages}{143--154}.
\newblock


\bibitem[Corbett et~al\mbox{.}(2013)]%
        {corbettSpanner2013}
\bibfield{author}{\bibinfo{person}{James~C. Corbett}, \bibinfo{person}{Jeffrey
  Dean}, \bibinfo{person}{Michael Epstein}, \bibinfo{person}{Andrew Fikes},
  \bibinfo{person}{Christopher Frost}, \bibinfo{person}{J.~J. Furman},
  \bibinfo{person}{Sanjay Ghemawat}, \bibinfo{person}{Andrey Gubarev},
  \bibinfo{person}{Christopher Heiser}, \bibinfo{person}{Peter Hochschild},
  \bibinfo{person}{Wilson Hsieh}, \bibinfo{person}{Sebastian Kanthak},
  \bibinfo{person}{Eugene Kogan}, \bibinfo{person}{Hongyi Li},
  \bibinfo{person}{Alexander Lloyd}, \bibinfo{person}{Sergey Melnik},
  \bibinfo{person}{David Mwaura}, \bibinfo{person}{David Nagle},
  \bibinfo{person}{Sean Quinlan}, \bibinfo{person}{Rajesh Rao},
  \bibinfo{person}{Lindsay Rolig}, \bibinfo{person}{Yasushi Saito},
  \bibinfo{person}{Michal Szymaniak}, \bibinfo{person}{Christopher Taylor},
  \bibinfo{person}{Ruth Wang}, {and} \bibinfo{person}{Dale Woodford}.}
  \bibinfo{year}{2013}\natexlab{}.
\newblock \showarticletitle{Spanner: {Google}'s {Globally} {Distributed}
  {Database}}.
\newblock \bibinfo{journal}{\emph{ACM Transactions on Computer Systems (TOCS
  '13)}} \bibinfo{volume}{31}, \bibinfo{number}{3} (\bibinfo{year}{2013}),
  \bibinfo{pages}{1--22}.
\newblock


\bibitem[Darold(2024)]%
        {noauthorPgbadgerNodate}
\bibfield{author}{\bibinfo{person}{Gilles Darold}.}
  \bibinfo{year}{2024}\natexlab{}.
\newblock \bibinfo{title}{{pgBadger}}.
\newblock
\urldef\tempurl%
\url{https://pgbadger.darold.net/}
\showURL{%
\tempurl}
\newblock
\shownote{Accessed: March 30, 2026}.


\bibitem[Dong et~al\mbox{.}(2021)]%
        {dongRocksdb2021}
\bibfield{author}{\bibinfo{person}{Siying Dong}, \bibinfo{person}{Andrew
  Kryczka}, \bibinfo{person}{Yanqin Jin}, {and} \bibinfo{person}{Michael
  Stumm}.} \bibinfo{year}{2021}\natexlab{}.
\newblock \showarticletitle{{RocksDB}: {Evolution} of {Development}
  {Priorities} in a {Key}-value {Store} {Serving} {Large}-scale
  {Applications}}.
\newblock \bibinfo{journal}{\emph{ACM Transactions on Storage. (TOS)}}
  \bibinfo{volume}{17}, \bibinfo{number}{4} (\bibinfo{year}{2021}),
  \bibinfo{pages}{1--32}.
\newblock


\bibitem[{eBPF Foundation}(2024)]%
        {noauthorEbpfNodate}
\bibfield{author}{\bibinfo{person}{{eBPF Foundation}}.}
  \bibinfo{year}{2024}\natexlab{}.
\newblock \bibinfo{title}{{eBPF} - {Introduction}, {Tutorials} \& {Community}
  {Resources}}.
\newblock
\urldef\tempurl%
\url{https://ebpf.io}
\showURL{%
\tempurl}
\newblock
\shownote{Accessed: March 30, 2026}.


\bibitem[Godard(2024)]%
        {noauthorIostat1Nodate}
\bibfield{author}{\bibinfo{person}{Sebastien Godard}.}
  \bibinfo{year}{2024}\natexlab{}.
\newblock \bibinfo{title}{iostat(1) - {Linux} man page}.
\newblock
\urldef\tempurl%
\url{https://linux.die.net/man/1/iostat}
\showURL{%
\tempurl}
\newblock
\shownote{Accessed: March 30, 2026}.


\bibitem[{Google}(2024a)]%
        {noauthorAlloydbNodate}
\bibfield{author}{\bibinfo{person}{{Google}}.}
  \bibinfo{year}{2024}\natexlab{a}.
\newblock \bibinfo{title}{{AlloyDB} for {PostgreSQL}}.
\newblock
\urldef\tempurl%
\url{https://cloud.google.com/products/alloydb}
\showURL{%
\tempurl}
\newblock
\shownote{Accessed: March 30, 2026}.


\bibitem[{Google}(2024b)]%
        {noauthorAlloydbNodate1}
\bibfield{author}{\bibinfo{person}{{Google}}.}
  \bibinfo{year}{2024}\natexlab{b}.
\newblock \bibinfo{title}{{AlloyDB} for {PostgreSQL} release notes}.
\newblock
\urldef\tempurl%
\url{https://cloud.google.com/alloydb/docs/release-notes}
\showURL{%
\tempurl}
\newblock
\shownote{Accessed: March 30, 2026}.


\bibitem[{Google}(2024c)]%
        {noauthorCloudNodate1}
\bibfield{author}{\bibinfo{person}{{Google}}.}
  \bibinfo{year}{2024}\natexlab{c}.
\newblock \bibinfo{title}{Cloud {SQL}}.
\newblock
\urldef\tempurl%
\url{https://cloud.google.com/sql}
\showURL{%
\tempurl}
\newblock
\shownote{Accessed: March 30, 2026}.


\bibitem[{Google}(2024d)]%
        {noauthorGoogleCalc}
\bibfield{author}{\bibinfo{person}{{Google}}.}
  \bibinfo{year}{2024}\natexlab{d}.
\newblock \bibinfo{title}{Google {Cloud} {Pricing} {Calculator}}.
\newblock
\urldef\tempurl%
\url{https://cloud.google.com/products/calculator?hl=en}
\showURL{%
\tempurl}
\newblock
\shownote{Accessed: March 30, 2026}.


\bibitem[{Google, Accenture}(2024)]%
        {accentureAlloydb2024}
\bibfield{author}{\bibinfo{person}{{Google, Accenture}}.}
  \bibinfo{year}{2024}\natexlab{}.
\newblock \bibinfo{title}{{AlloyDB}, an {Option} for {Enterprise} {Workloads}?}
\newblock
\urldef\tempurl%
\url{https://www.accenture.com/content/dam/accenture/final/accenture-com/document-2/Accenture-AlloyDB-For-Enterprise-Workloads.pdf}
\showURL{%
\tempurl}
\newblock
\shownote{Accessed: March 30, 2026}.


\bibitem[Gregg(2020)]%
        {Gregg2020}
\bibfield{author}{\bibinfo{person}{Brendan Gregg}.}
  \bibinfo{year}{2020}\natexlab{}.
\newblock \bibinfo{booktitle}{\emph{Systems Performance: Enterprise and the
  Cloud} (\bibinfo{edition}{2nd} ed.)}.
\newblock \bibinfo{publisher}{Addison-Wesley Professional},
  \bibinfo{address}{Boston, MA}.
\newblock
\showISBNx{978-0136820154}


\bibitem[Gulati et~al\mbox{.}(2007)]%
        {gulatiDClock2007}
\bibfield{author}{\bibinfo{person}{Ajay Gulati}, \bibinfo{person}{Arif
  Merchant}, {and} \bibinfo{person}{Peter Varman}.}
  \bibinfo{year}{2007}\natexlab{}.
\newblock \showarticletitle{d-clock: distributed {QoS} in heterogeneous
  resource environments}. In \bibinfo{booktitle}{\emph{Proceedings of the {ACM}
  Symposium on {Principles} of Distributed Computing. (PODC '07)}}.
  \bibinfo{address}{Portland, OR, USA}, \bibinfo{pages}{330--331}.
\newblock


\bibitem[Gulati et~al\mbox{.}(2010)]%
        {gulatiMclock2010}
\bibfield{author}{\bibinfo{person}{Ajay Gulati}, \bibinfo{person}{Arif
  Merchant}, {and} \bibinfo{person}{Peter~J. Varman}.}
  \bibinfo{year}{2010}\natexlab{}.
\newblock \showarticletitle{{mClock}: handling throughput variability for
  hypervisor {IO} scheduling}. In \bibinfo{booktitle}{\emph{Proceedings of the
  {USENIX} Conference on {Operating} Systems Design and Implementation. (OSDI
  '10)}}. \bibinfo{publisher}{USENIX Association}, \bibinfo{address}{Vancouver,
  BC, Canada}, \bibinfo{pages}{437--450}.
\newblock


\bibitem[{HammerDB}(2024)]%
        {noauthorHammerdbNodate}
\bibfield{author}{\bibinfo{person}{{HammerDB}}.}
  \bibinfo{year}{2024}\natexlab{}.
\newblock \bibinfo{title}{{HammerDB}}.
\newblock
\urldef\tempurl%
\url{https://www.hammerdb.com/}
\showURL{%
\tempurl}
\newblock
\shownote{Accessed: March 30, 2026}.


\bibitem[{HashiCorp}(2024)]%
        {hashicorpTerraform}
\bibfield{author}{\bibinfo{person}{{HashiCorp}}.}
  \bibinfo{year}{2024}\natexlab{}.
\newblock \bibinfo{title}{Terraform by {HashiCorp}}.
\newblock
\urldef\tempurl%
\url{https://www.terraform.io/}
\showURL{%
\tempurl}
\newblock
\shownote{Accessed: March 30, 2026}.


\bibitem[Huang et~al\mbox{.}(2020)]%
        {huangTidb2020}
\bibfield{author}{\bibinfo{person}{Dongxu Huang}, \bibinfo{person}{Qi Liu},
  \bibinfo{person}{Qiu Cui}, \bibinfo{person}{Zhuhe Fang},
  \bibinfo{person}{Xiaoyu Ma}, \bibinfo{person}{Fei Xu}, \bibinfo{person}{Li
  Shen}, \bibinfo{person}{Liu Tang}, \bibinfo{person}{Yuxing Zhou},
  \bibinfo{person}{Menglong Huang}, \bibinfo{person}{Wan Wei},
  \bibinfo{person}{Cong Liu}, \bibinfo{person}{Jian Zhang},
  \bibinfo{person}{Jianjun Li}, \bibinfo{person}{Xuelian Wu},
  \bibinfo{person}{Lingyu Song}, \bibinfo{person}{Ruoxi Sun},
  \bibinfo{person}{Shuaipeng Yu}, \bibinfo{person}{Lei Zhao},
  \bibinfo{person}{Nicholas Cameron}, \bibinfo{person}{Liquan Pei}, {and}
  \bibinfo{person}{Xin Tang}.} \bibinfo{year}{2020}\natexlab{}.
\newblock \showarticletitle{{TiDB}: a {Raft}-based {HTAP} database}.
\newblock \bibinfo{journal}{\emph{Proceedings of the VLDB Endowment. (PVLDB)}}
  \bibinfo{volume}{13}, \bibinfo{number}{12} (\bibinfo{year}{2020}),
  \bibinfo{pages}{3072--3084}.
\newblock


\bibitem[{IO Visor Project}(2024)]%
        {noauthorBccNodate}
\bibfield{author}{\bibinfo{person}{{IO Visor Project}}.}
  \bibinfo{year}{2024}\natexlab{}.
\newblock \bibinfo{title}{bcc}.
\newblock
\urldef\tempurl%
\url{https://github.com/iovisor/bcc}
\showURL{%
\tempurl}
\newblock
\shownote{Accessed: March 30, 2026}.


\bibitem[Kim et~al\mbox{.}(2023)]%
        {kimRevitalizing2023}
\bibfield{author}{\bibinfo{person}{Jongseok Kim}, \bibinfo{person}{Chanu Yu},
  {and} \bibinfo{person}{Euiseong Seo}.} \bibinfo{year}{2023}\natexlab{}.
\newblock \showarticletitle{Revitalizing {Buffered} {I}/{O}: {Optimizing}
  {Page} {Reclaim} and {I}/{O} {Throttling}}. In \bibinfo{booktitle}{\emph{2023
  {IEEE} 41st {International} {Conference} on {Computer} {Design}. (ICCD)}}.
  \bibinfo{publisher}{IEEE}, \bibinfo{address}{Washington, DC, USA},
  \bibinfo{pages}{475--482}.
\newblock


\bibitem[Lamport(1998)]%
        {lamportPartTime1998}
\bibfield{author}{\bibinfo{person}{Leslie Lamport}.}
  \bibinfo{year}{1998}\natexlab{}.
\newblock \showarticletitle{The part-time parliament}.
\newblock \bibinfo{journal}{\emph{ACM Trans. Comput. Syst.}}
  \bibinfo{volume}{16}, \bibinfo{number}{2} (\bibinfo{year}{1998}),
  \bibinfo{pages}{133--169}.
\newblock


\bibitem[McIlroy et~al\mbox{.}(1978)]%
        {macilroy}
\bibfield{author}{\bibinfo{person}{M.D. McIlroy}, \bibinfo{person}{E.N.
  Pinson}, {and} \bibinfo{person}{B.A. Tague}.}
  \bibinfo{year}{1978}\natexlab{}.
\newblock \showarticletitle{UNIX Time-Sharing System: Foreword}.
\newblock \bibinfo{journal}{\emph{Bell System Technical Journal}}
  \bibinfo{volume}{57}, \bibinfo{number}{6} (\bibinfo{year}{1978}),
  \bibinfo{pages}{1899--1904}.
\newblock
\newblock
\shownote{Part 2}.


\bibitem[Mohan et~al\mbox{.}(1992)]%
        {mohanAries1992}
\bibfield{author}{\bibinfo{person}{C. Mohan}, \bibinfo{person}{Don Haderle},
  \bibinfo{person}{Bruce Lindsay}, \bibinfo{person}{Hamid Pirahesh}, {and}
  \bibinfo{person}{Peter Schwarz}.} \bibinfo{year}{1992}\natexlab{}.
\newblock \showarticletitle{{ARIES}: a transaction recovery method supporting
  fine-granularity locking and partial rollbacks using write-ahead logging}.
\newblock \bibinfo{journal}{\emph{ACM Transactions. Database Systems.}}
  \bibinfo{volume}{17}, \bibinfo{number}{1} (\bibinfo{year}{1992}),
  \bibinfo{pages}{94--162}.
\newblock


\bibitem[Mozaffari et~al\mbox{.}(2024)]%
        {mozaffariSelfTuning2024}
\bibfield{author}{\bibinfo{person}{Maryam Mozaffari}, \bibinfo{person}{Anton
  Dign\"{o}s}, \bibinfo{person}{Johann Gamper}, {and} \bibinfo{person}{Uta
  St\"{o}rl}.} \bibinfo{year}{2024}\natexlab{}.
\newblock \showarticletitle{Self-tuning {Database} {Systems}: {A} {Systematic}
  {Literature} {Review} of {Automatic} {Database} {Schema} {Design} and
  {Tuning}}.
\newblock \bibinfo{journal}{\emph{ACM Computing Surveys.}}
  \bibinfo{volume}{56}, \bibinfo{number}{11} (\bibinfo{year}{2024}),
  \bibinfo{pages}{277:1--277:37}.
\newblock


\bibitem[{Neon}(2024)]%
        {noauthorNeonNodate}
\bibfield{author}{\bibinfo{person}{{Neon}}.} \bibinfo{year}{2024}\natexlab{}.
\newblock \bibinfo{title}{Neon {Serverless} {Postgres}}.
\newblock
\urldef\tempurl%
\url{https://neon.tech}
\showURL{%
\tempurl}
\newblock
\shownote{Accessed: March 30, 2026}.


\bibitem[Ongaro and Ousterhout(2014)]%
        {ongaroSearch2014}
\bibfield{author}{\bibinfo{person}{Diego Ongaro} {and} \bibinfo{person}{John
  Ousterhout}.} \bibinfo{year}{2014}\natexlab{}.
\newblock \showarticletitle{In search of an understandable consensus
  algorithm}. In \bibinfo{booktitle}{\emph{Proceedings of the {USENIX}
  conference on {USENIX} {Annual} {Technical} {Conference}. (USENIX ATC '14)}}.
  \bibinfo{publisher}{USENIX Association}, \bibinfo{address}{Philadelphia, PA},
  \bibinfo{pages}{305--320}.
\newblock


\bibitem[{Oracle}(2024a)]%
        {noauthorMysqlNodate}
\bibfield{author}{\bibinfo{person}{{Oracle}}.}
  \bibinfo{year}{2024}\natexlab{a}.
\newblock \bibinfo{title}{{MySQL}}.
\newblock
\urldef\tempurl%
\url{https://www.mysql.com/}
\showURL{%
\tempurl}
\newblock
\shownote{Accessed: March 30, 2026}.


\bibitem[{Oracle}(2024b)]%
        {noauthorOracleNodate}
\bibfield{author}{\bibinfo{person}{{Oracle}}.}
  \bibinfo{year}{2024}\natexlab{b}.
\newblock \bibinfo{title}{Oracle {Autonomous} {Transaction} {Processing}}.
\newblock
\urldef\tempurl%
\url{https://www.oracle.com/autonomous-database/autonomous-transaction-processing/}
\showURL{%
\tempurl}
\newblock
\shownote{Accessed: March 30, 2026}.


\bibitem[{Oracle}(2024c)]%
        {noauthorOtnNodate}
\bibfield{author}{\bibinfo{person}{{Oracle}}.}
  \bibinfo{year}{2024}\natexlab{c}.
\newblock \bibinfo{title}{{OTN} {Development} and {Distribution} {License}
  {Terms}}.
\newblock
\urldef\tempurl%
\url{https://www.oracle.com/downloads/licenses/instant-client-lic.html}
\showURL{%
\tempurl}
\newblock
\shownote{Accessed: March 30, 2026}.


\bibitem[{OSSC-DB}(2024)]%
        {noauthorPgStatsinfoNodate}
\bibfield{author}{\bibinfo{person}{{OSSC-DB}}.}
  \bibinfo{year}{2024}\natexlab{}.
\newblock \bibinfo{title}{pg\_statsinfo}.
\newblock
\urldef\tempurl%
\url{https://github.com/ossc-db/pg_statsinfo}
\showURL{%
\tempurl}
\newblock
\shownote{Accessed: March 30, 2026}.


\bibitem[Pang and Wang(2024)]%
        {pangUnderstanding2024}
\bibfield{author}{\bibinfo{person}{Xi Pang} {and} \bibinfo{person}{Jianguo
  Wang}.} \bibinfo{year}{2024}\natexlab{}.
\newblock \showarticletitle{Understanding the {Performance} {Implications} of
  the {Design} {Principles} in {Storage}-{Disaggregated} {Databases}}. In
  \bibinfo{booktitle}{\emph{Proceedings of the ACM International Conference on
  Management of Data. (SIGMOD '24)}}. \bibinfo{publisher}{ACM},
  \bibinfo{address}{New York, NY, USA}, \bibinfo{pages}{180:1--180:26}.
\newblock


\bibitem[Park et~al\mbox{.}(2020)]%
        {park2020iops}
\bibfield{author}{\bibinfo{person}{Hojin Park}, \bibinfo{person}{Gregory~R.
  Ganger}, {and} \bibinfo{person}{George Amvrosiadis}.}
  \bibinfo{year}{2020}\natexlab{}.
\newblock \showarticletitle{More {IOPS} for Less: Exploiting Burstable Storage
  in Public Clouds}. In \bibinfo{booktitle}{\emph{12th USENIX Workshop on Hot
  Topics in Cloud Computing (HotCloud '20)}}. \bibinfo{publisher}{USENIX
  Association}, \bibinfo{address}{Virtual}.
\newblock


\bibitem[{PostgreSQL Global Development Group}(2024)]%
        {groupPostgresql2024}
\bibfield{author}{\bibinfo{person}{{PostgreSQL Global Development Group}}.}
  \bibinfo{year}{2024}\natexlab{}.
\newblock \bibinfo{title}{{PostgreSQL}}.
\newblock
\urldef\tempurl%
\url{https://www.postgresql.org/}
\showURL{%
\tempurl}
\newblock
\shownote{Accessed: March 30, 2026}.


\bibitem[{Pulumi}(2024)]%
        {pulumiPulumi}
\bibfield{author}{\bibinfo{person}{{Pulumi}}.} \bibinfo{year}{2024}\natexlab{}.
\newblock \bibinfo{title}{Pulumi - Infrastructure as Code}.
\newblock
\urldef\tempurl%
\url{https://www.pulumi.com/}
\showURL{%
\tempurl}
\newblock
\shownote{Accessed: March 30, 2026}.


\bibitem[Qian et~al\mbox{.}(2024)]%
        {qianCombining2024}
\bibfield{author}{\bibinfo{person}{Yingjin Qian},
  \bibinfo{person}{Marc-Andr\'{e} Vef}, \bibinfo{person}{Patrick Farrell},
  \bibinfo{person}{Andreas Dilger}, \bibinfo{person}{Xi Li},
  \bibinfo{person}{Shuichi Ihara}, \bibinfo{person}{Yinjin Fu},
  \bibinfo{person}{Wei Xue}, {and} \bibinfo{person}{Andr\'{e} Brinkmann}.}
  \bibinfo{year}{2024}\natexlab{}.
\newblock \showarticletitle{Combining buffered {I}/{O} and direct {I}/{O} in
  distributed file systems}. In \bibinfo{booktitle}{\emph{Proceedings of the
  {USENIX} {Conference} on {File} and {Storage} {Technologies}. (FAST '24)}}.
  \bibinfo{publisher}{USENIX Association}, \bibinfo{address}{Santa Clara, CA},
  \bibinfo{pages}{17--34}.
\newblock


\bibitem[{Red Hat}(2024)]%
        {redhatAnsible}
\bibfield{author}{\bibinfo{person}{{Red Hat}}.}
  \bibinfo{year}{2024}\natexlab{}.
\newblock \bibinfo{title}{Ansible by {Red Hat}}.
\newblock
\urldef\tempurl%
\url{https://www.ansible.com/}
\showURL{%
\tempurl}
\newblock
\shownote{Accessed: March 30, 2026}.


\bibitem[Robin and Harshita(2025)]%
        {gartner2025dbmsmarketshare}
\bibfield{author}{\bibinfo{person}{Schumacher Robin} {and}
  \bibinfo{person}{Chibber Harshita}.} \bibinfo{year}{2025}\natexlab{}.
\newblock \bibinfo{booktitle}{\emph{Market Share: Database Management Systems,
  Worldwide, 2024}}.
\newblock \bibinfo{type}{{T}echnical {R}eport}. \bibinfo{institution}{Gartner,
  Inc.}
\newblock
\newblock
\shownote{Document ID: G00806029.
  \url{https://www.gartner.com/en/documents/6494271}. Accessed: March 30,
  2026}.


\bibitem[Taft et~al\mbox{.}(2020)]%
        {taftCockroachdb2020}
\bibfield{author}{\bibinfo{person}{Rebecca Taft}, \bibinfo{person}{Irfan
  Sharif}, \bibinfo{person}{Andrei Matei}, \bibinfo{person}{Nathan
  VanBenschoten}, \bibinfo{person}{Jordan Lewis}, \bibinfo{person}{Tobias
  Grieger}, \bibinfo{person}{Kai Niemi}, \bibinfo{person}{Andy Woods},
  \bibinfo{person}{Anne Birzin}, \bibinfo{person}{Raphael Poss},
  \bibinfo{person}{Paul Bardea}, \bibinfo{person}{Amruta Ranade},
  \bibinfo{person}{Ben Darnell}, \bibinfo{person}{Bram Gruneir},
  \bibinfo{person}{Justin Jaffray}, \bibinfo{person}{Lucy Zhang}, {and}
  \bibinfo{person}{Peter Mattis}.} \bibinfo{year}{2020}\natexlab{}.
\newblock \showarticletitle{{CockroachDB}: {The} {Resilient}
  {Geo}-{Distributed} {SQL} {Database}}. In
  \bibinfo{booktitle}{\emph{Proceedings of the ACM International Conference on
  Management of Data. (SIGMOD '20)}}. \bibinfo{publisher}{ACM},
  \bibinfo{address}{New York, NY, USA}, \bibinfo{pages}{1493--1509}.
\newblock


\bibitem[Torvalds(2025)]%
        {torvaldsTorvaldslinux2025}
\bibfield{author}{\bibinfo{person}{Linus Torvalds}.}
  \bibinfo{year}{2025}\natexlab{}.
\newblock \bibinfo{title}{torvalds/linux}.
\newblock
\urldef\tempurl%
\url{https://github.com/torvalds/linux}
\showURL{%
\tempurl}
\newblock
\shownote{Accessed: March 30, 2026}.


\bibitem[{Transaction Processing Performance Council}(2024)]%
        {noauthorTpcCNodate}
\bibfield{author}{\bibinfo{person}{{Transaction Processing Performance
  Council}}.} \bibinfo{year}{2024}\natexlab{}.
\newblock \bibinfo{title}{{TPC}-{C} {Homepage}}.
\newblock
\urldef\tempurl%
\url{https://www.tpc.org/tpcc/}
\showURL{%
\tempurl}
\newblock
\shownote{Accessed: March 30, 2026}.


\bibitem[Verbitski et~al\mbox{.}(2017)]%
        {verbitskiAmazon2017}
\bibfield{author}{\bibinfo{person}{Alexandre Verbitski},
  \bibinfo{person}{Anurag Gupta}, \bibinfo{person}{Debanjan Saha},
  \bibinfo{person}{Murali Brahmadesam}, \bibinfo{person}{Kamal Gupta},
  \bibinfo{person}{Raman Mittal}, \bibinfo{person}{Sailesh Krishnamurthy},
  \bibinfo{person}{Sandor Maurice}, \bibinfo{person}{Tengiz Kharatishvili},
  {and} \bibinfo{person}{Xiaofeng Bao}.} \bibinfo{year}{2017}\natexlab{}.
\newblock \showarticletitle{Amazon {Aurora}: {Design} {Considerations} for
  {High} {Throughput} {Cloud}-{Native} {Relational} {Databases}}. In
  \bibinfo{booktitle}{\emph{Proceedings of the ACM International Conference on
  Management of Data. (SIGMOD '17)}}. \bibinfo{publisher}{ACM},
  \bibinfo{address}{New York, NY, USA}, \bibinfo{pages}{1041--1052}.
\newblock


\bibitem[Warfield(2023)]%
        {warfieldAmazon2023}
\bibfield{author}{\bibinfo{person}{Andy Warfield}.}
  \bibinfo{year}{2023}\natexlab{}.
\newblock \showarticletitle{Amazon {S3}}. In \bibinfo{booktitle}{\emph{21st
  USENIX Conference on File and Storage Technologies (FAST 23)}}.
  \bibinfo{publisher}{USENIX Association}, \bibinfo{address}{Santa Clara, CA}.
\newblock
\newblock
\shownote{Place: Santa Clara, CA. Publisher: USENIX Association}.


\bibitem[Weil et~al\mbox{.}(2006)]%
        {weilCeph2006}
\bibfield{author}{\bibinfo{person}{Sage~A. Weil}, \bibinfo{person}{Scott~A.
  Brandt}, \bibinfo{person}{Ethan~L. Miller}, \bibinfo{person}{Darrell D.~E.
  Long}, {and} \bibinfo{person}{Carlos Maltzahn}.}
  \bibinfo{year}{2006}\natexlab{}.
\newblock \showarticletitle{Ceph: a scalable, high-performance distributed file
  system}. In \bibinfo{booktitle}{\emph{Proceedings of {Operating} Systems
  Design and Implementation. (OSDI '06)}}. \bibinfo{publisher}{USENIX
  Association}, \bibinfo{address}{Seattle, WA}, \bibinfo{pages}{307--320}.
\newblock


\bibitem[Wu(2009)]%
        {noauthorWritebackNodate}
\bibfield{author}{\bibinfo{person}{Fengguang Wu}.}
  \bibinfo{year}{2009}\natexlab{}.
\newblock \bibinfo{title}{writeback: switch to per-bdi threads for flushing
  data \textperiodcentered{} torvalds/linux@03ba378}.
\newblock
\urldef\tempurl%
\url{https://github.com/torvalds/linux/commit/03ba3782e8dcc5b0e1efe440d33084f066e38cae}
\showURL{%
\tempurl}
\newblock
\shownote{Accessed: March 30, 2026}.


\bibitem[Wu(2012)]%
        {fengguangFengguangNodate}
\bibfield{author}{\bibinfo{person}{Fengguang Wu}.}
  \bibinfo{year}{2012}\natexlab{}.
\newblock \bibinfo{title}{Fengguang, {W}. {IO}-less {Dirty} {Throttling}}.
\newblock
\urldef\tempurl%
\url{https://events.static.linuxfound.org/images/stories/pdf/lcjp2012_wu.pdf}
\showURL{%
\tempurl}
\newblock
\shownote{Accessed: March 30, 2026}.


\bibitem[Yang et~al\mbox{.}(2024)]%
        {yangPolardbMp2024}
\bibfield{author}{\bibinfo{person}{Xinjun Yang}, \bibinfo{person}{Yingqiang
  Zhang}, \bibinfo{person}{Hao Chen}, \bibinfo{person}{Feifei Li},
  \bibinfo{person}{Bo Wang}, \bibinfo{person}{Jing Fang},
  \bibinfo{person}{Chuan Sun}, {and} \bibinfo{person}{Yuhui Wang}.}
  \bibinfo{year}{2024}\natexlab{}.
\newblock \showarticletitle{{PolarDB}-{MP}: {A} {Multi}-{Primary}
  {Cloud}-{Native} {Database} via {Disaggregated} {Shared} {Memory}}. In
  \bibinfo{booktitle}{\emph{Proceedings of the ACM International Conference on
  Management of Data. (SIGMOD '24)}}. \bibinfo{publisher}{ACM},
  \bibinfo{address}{New York, NY, USA}, \bibinfo{pages}{295--308}.
\newblock


\bibitem[Yang et~al\mbox{.}(2022)]%
        {yangOceanbase2022}
\bibfield{author}{\bibinfo{person}{Zhenkun Yang}, \bibinfo{person}{Chuanhui
  Yang}, \bibinfo{person}{Fusheng Han}, \bibinfo{person}{Mingqiang Zhuang},
  \bibinfo{person}{Bing Yang}, \bibinfo{person}{Zhifeng Yang},
  \bibinfo{person}{Xiaojun Cheng}, \bibinfo{person}{Yuzhong Zhao},
  \bibinfo{person}{Wenhui Shi}, \bibinfo{person}{Huafeng Xi},
  \bibinfo{person}{Huang Yu}, \bibinfo{person}{Bin Liu}, \bibinfo{person}{Yi
  Pan}, \bibinfo{person}{Boxue Yin}, \bibinfo{person}{Junquan Chen}, {and}
  \bibinfo{person}{Quanqing Xu}.} \bibinfo{year}{2022}\natexlab{}.
\newblock \showarticletitle{{OceanBase}: a 707 million {tpmC} distributed
  relational database system}.
\newblock \bibinfo{journal}{\emph{Proceedings of the VLDB Endowment. (PVLDB)}}
  \bibinfo{volume}{15}, \bibinfo{number}{12} (\bibinfo{year}{2022}),
  \bibinfo{pages}{3385--3397}.
\newblock


\bibitem[{Yugabyte}(2024)]%
        {noauthorDistributedNodate}
\bibfield{author}{\bibinfo{person}{{Yugabyte}}.}
  \bibinfo{year}{2024}\natexlab{}.
\newblock \bibinfo{title}{Distributed {PostgreSQL} for {Modern} {Apps}}.
\newblock
\urldef\tempurl%
\url{https://www.yugabyte.com/}
\showURL{%
\tempurl}
\newblock
\shownote{Accessed: March 30, 2026}.


\end{thebibliography}
}

\end{document}